\documentclass[
10pt,
]{nature}

\usepackage{graphicx}

\usepackage{microtype}
\usepackage{booktabs}
\usepackage{soul}
\usepackage[para]{threeparttable}
 \usepackage{lineno}
\usepackage[english]{babel}
\usepackage[utf8]{inputenc}
\usepackage[pdftex]{hyperref, color}
\usepackage[figure, figure*]{hypcap}

\usepackage{upgreek}
\usepackage{lmodern}
\usepackage{wrapfig}
\usepackage{dcolumn}
\usepackage{bm}

\usepackage{bbold}
\usepackage[T1]{fontenc}
\usepackage{color,bm, braket}
\usepackage[emulate = units]{siunitx}
\usepackage{xr}
\usepackage{xcolor}

\newcommand {\e}{\mathrm{e}}
\newcommand {\mo}{MoS$_2$}
\newcommand{\nG}{n-doped Gr}
\newcommand{\pG}{p-doped Gr}

	\DeclareGraphicsExtensions{.pdf}

\begin{document}
\title{Metal-insulator transition in monolayer \mo{} via contactless chemical doping}
\maketitle
\author{Camiel van Efferen$^{1,5}*$,}
\author{Clifford Murray$^{1,5}$,}
\author{Jeison Fischer$^1$,} 
\author{Carsten Busse$^{2,3}$,}
\author{Hannu-Pekka Komsa$^4$,}
\author{Thomas Michely$^1$,}
\author{Wouter Jolie$^{1}$} 
\begin{affiliations}
\item II. Physikalisches Institut, Universit\"{a}t zu K\"{o}ln, Z\"{u}lpicher Stra\ss e 77, 50937 K\"{o}ln, Germany
\item Present address: Department Physik, Universität Siegen, Walter-Flex-Str. 3, 57068 Siegen, Germany
\item Institut f{\"ur} Materialphysik, Westf\"{a}lische Wilhelms-Universit\"{a}t M\"{u}nster, Wilhelm-Klemm-Stra{\ss}e 10, 48149 M\"{u}nster, Germany
\item Faculty of Information Technology and Electrical Engineering, University of Oulu, Pentti Kaiteran katu 1, 90014 Oulu, Finland
\item CvE and CM contributed equally to this work
\end{affiliations}

%\clearpage	

	\begin{abstract}
Much effort has been made to modify the properties of transition metal dichalcogenide layers \textit{via} their environment as a route to new functionalization. However, it remains a challenge to induce large electronic changes without chemically altering the layer or compromising its two-dimensionality. Here, a non-invasive technique is used to shift the chemical potential of monolayer \mo{} through p- and n-type doping of graphene (Gr), which remains a well-decoupled 2D substrate. With the intercalation of oxygen (O) under Gr, a nearly rigid Fermi level shift of \SI{0.45}{\eV} in \mo{} is demonstrated, whereas the intercalation of europium (Eu) induces a metal-insulator transition in \mo, accompanied by a giant band gap reduction of \SI{0.67}{\eV}. Additionally, the effect of the substrate charge on 1D states within \mo{} mirror-twin boundaries (MTBs) is explored. It is found that the 1D nature of the MTB states is not compromised, even when \mo{} is made metallic. Furthermore, with the periodicity of the 1D states dependent on substrate-induced charging and depletion, the boundaries serve as chemical potential sensors functional up to room temperature.

	\end{abstract}

%\clearpage

	\begin{figure*}
		\centering
		\includegraphics[width=0.9\textwidth]{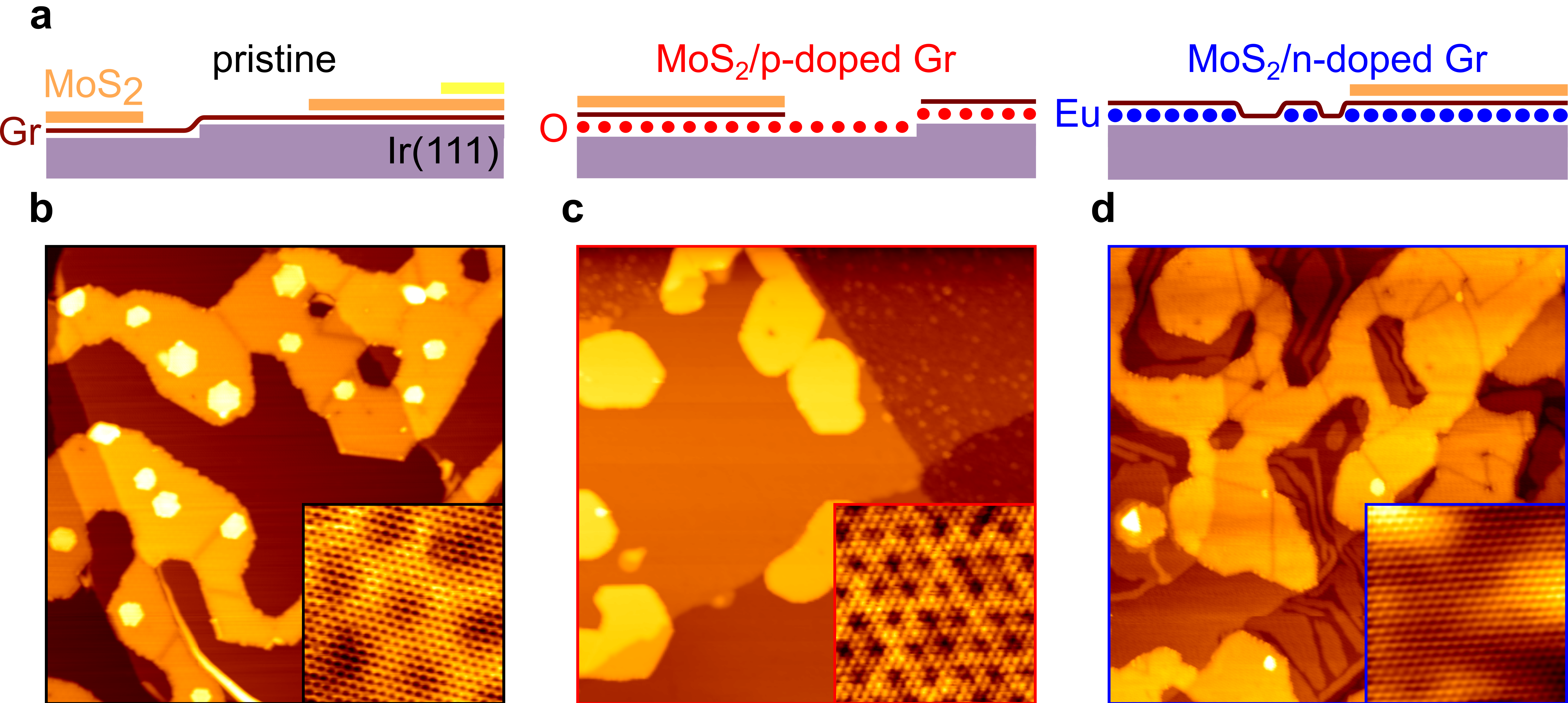}
		\caption{\footnotesize \textbf{a}~Side view schematics visualizing the method of achieving large chemical potential shifts of \mo{} on Gr/Ir(111). Left: pristine \mo. Middle: \mo{} on p-doped Gr through post-growth O intercalation under Gr. Right: \mo{} on n-doped Gr through pre-growth intercalation of Eu under Gr. \textbf{b-d}~Constant current STM overview topographs of: \textbf{b}~\mo/Gr/Ir(111); \textbf{c}~\mo/Gr/O/Ir(111) and \textbf{d}~\mo/Gr/Eu/Ir(111). \textbf{b}~Sample largely covered by ML \mo, with small hexagonal bilayer islands. Two Ir(111) step edges and, at the bottom, a Gr wrinkle are visible. The atomic resolution inset shows the moir\'e of Gr on Ir(111) imaged through the \mo. \textbf{c}~Sample sparsely covered by ML \mo; no bilayers have formed. On the right, Ir(111) is not covered by Gr. The atomic resolution inset shows that the appearance of a \mo/Gr moir\'e. \textbf{d}~ Sample largely covered by ML \mo, with close to no bilayers present. Gr appears in two phases, bare (dark) and Eu-intercalated (brighter). As is highlighted by the inset, no Gr/Ir(111) moir\'e appears in the \mo{} layer.  
	STM parameters ($V$, $I$): \textbf{b}~\SI{1.5}{\V}, \SI{0.01}{\nA}; \textbf{c}~\SI{3.5}{\V}, \SI{0.01}{\nA}; \textbf{d}~\SI{1.0}{\V}, \SI{0.10}{\nA}; \textbf{b-d}~image size \SI[parse-numbers=false]{150 \times 150}{\nano\meter\squared}. Insets: \textbf{b}~\SI{0.5}{\V}, \SI{0.58}{\nA}; \textbf{c}~\SI{1.0}{\V}, \SI{0.10}{\nA}; \textbf{d}~\SI{1.0}{\V}, \SI{0.10}{\nA}; \textbf{b-d}~image size \SI[parse-numbers=false]{6.3 \times 6.3}{\nano\meter\squared}.
			\label{fig1}}
	\end{figure*}

Metallic transition metal dichalcogenide (TMDC) monolayers show a plethora of many-body phenomena, such as charge density wave order~\cite{Xi2015}, Mott insulating states~\cite{Chen2020a} or superconductivity~\cite{Ugeda2016}. Understanding these correlated phases, as well as their dependence on external electric or magnetic fields, allows the creation of devices with tunable phase transitions~\cite{Li2016c}. While intrinsic TMDC metals are being extensively investigated, less is known about the metallic properties of intrinsic TMDC semiconductors such as \mo~and WS$_2$, mainly due to experimental difficulties in shifting the chemical potential close to the conduction or valence band.

Monolayer \mo~and WS$_2$~have electronic band gaps of more than \SI{2.5}{\eV}~\cite{Murray2019, Schuler2019b}. Hence, back-gated devices are typically unable to bring the chemical potential close to the onset of the conduction or valence band before breakdown occurs~\cite{Nguyen2019,Qiu2019}. Larger shifts are obtained with ionic-liquid-gating. With this method, transport measurements showed that \mo~and WS$_2$~become superconducting when the chemical potential lies within the conduction band~\cite{Ye2012,Lu2015a,Saito2016a,Saito2016,Lu2018,Wang2020c}. However, ionic-liquid-gating suffers from charge inhomogeneities~\cite{Ren2015}. At low temperatures this effect is exacerbated due to freezing of the liquid~\cite{Jo2015, Costanzo2016}. The liquid also hampers complementary measurements using surface sensitive techniques. As a consequence, many open questions remain on the nature of the superconducting dome in \mo~\cite{Ye2012,Lu2015a,Saito2016a,Fu2017,Piatti2018} and WS$_2$~\cite{Lu2018}, the origin of the finite density of states within the superconducting gap \cite{Costanzo2018}, and on the properties of the (quasi-) metallic phase in between the insulating and superconducting phases~\cite{Lu2018}.

Another method to obtain large shifts of the chemical potential is the introduction of foreign species. This is done through adsorption, elemental substitution or \textit{via} intercalation between the 2D sheet and its substrate~\cite{Helveg2000a,Alidoust2014,Zhang2014d,Kang2017,Katoch2018,Liu2020c}. But the direct contact with the TMDC results in, like in the case of ionic-gating, an entangled effect of charge redistribution and chemical hybridization \cite{Shao2019}, leading to non-universal, element-specific properties. Hence, an efficient method able to tune the chemical potential of TMDCs by large amounts without chemically altering it or leaving the surface inaccessible would be highly desirable. Such a method would make it possible to monitor the effect of charge on the band structure of TMDC semiconductors \cite{Wang2020c}, ultimately enabling access to novel phases of matter, such as topological superconductivity \cite{Hsu2017}.

Here, we introduce contactless chemical doping as a method to induce large shifts in the chemical potential of \mo, without disturbing its chemical environment. We chemically dope a Gr layer from below in order to shift the chemical potential of \mo{} monolayer grown on top \textit{in situ}. The doping is done through the use of intercalants, taking Eu and O as electron donor and acceptor respectively, to create high quality \mo/Gr/(Eu or O)/Ir(111) layered configurations. For both intercalants, it has been found that they are able to cause considerable shifts in the Dirac point of Gr (as far as $E_{\text{D}} = \SI{-1.38}{\eV}$ for Eu and up to $E_{\text{D}} =+\SI{0.68}{\eV}$ for O), without strong hybridization~\cite{Schumacher2013a,Larciprete2012, Granas2012, Jolie2014}. In this way, the chemical potential within the \mo~band structure can be controlled through the doping level of Gr and shifted far beyond what is possible in traditional back-gate device setups, without hybridization or loss of surface accessibility and quality. The \mo/doped Gr samples are analyzed experimentally with scanning tunneling microscopy (STM) and spectroscopy (STS) and theoretically with \textit{ab initio} density functional theory (DFT) calculations. Additional attention is given to the effect of the Gr doping on the 1D states within mirror-twin boundaries (MTBs) in \mo{}, which can be filled and depleted \textit{via} the substrate without compromising their 1D nature. Thereby, they can be used as chemical potential sensors.

%%%%%%%%%%%%%%%%%%%%%%%%%%%%%%%%%%%%%%%%%%%%%%%%%%%%%%%%%%%%%%%%%%%%%%%%%%%%%%%%%%%%%%%%%%%%%%%%%%%%%%%%%%

	\section*{Results}
	
	\subsection{Metal-insulator transition in \mo}\label{secSubsMacro}
	
Our method is illustrated in Fig.~\ref{fig1}a. The initial heterostructure consists of three different materials: the 2D semiconductor \mo, the 2D semi-metal Gr, with the Dirac point close to Fermi level $E_{\text{F}}$, and the metallic Ir(111). Since foreign elements bind much more strongly to Ir(111) than to the van der Waals materials Gr and \mo, this arrangement enables us to intercalate these between Gr and Ir(111), while preserving the chemical environment of \mo. Here, the concept is demonstrated for the case of Eu and O. The applicability of our method to other 2D semiconductors is demonstrated in the SI (Fig.~S1).

Intercalating an electron donor such as Eu leads to an upward shift of the chemical potential in Gr \textit{via} charge transfer. Similarly, the intercalation of an electron acceptor such as O causes the chemical potential of Gr to shift down in energy. These changes in the Gr substrate cause comparable shifts in the chemical potential of the top layer \textit{via} workfunction alignment and/or charge transfer. The precise mechanisms by which the \mo{} chemical potential is shifted for each intercalant will be discussed in detail below. In the following, we refer to \mo/Gr/Eu/Ir(111) [\mo/Gr/O/Ir(111)] as \mo~on n-doped [p-doped] Gr, while \mo/Gr/Ir(111) is refered to as pristine.

The STM topograph of Fig.~\ref{fig1}b displays the pristine sample as a reference. Islands of monolayer (ML) \mo{} extend across the Gr/Ir(111) substrate, including a few Ir(111) step edges. The typical apparent height of the ML islands is \SI{0.65}{\nm} when tunneling into the \mo{} conduction band. Small BL-\mo{} islands are seen as well. Apart from MTBs, which appear as dark, straight lines at high bias voltages, defects are largly absent. In the inset, an atomic resolution topograph of \mo{} shows that the moir\'e of Gr/Ir(111) is visible through the \mo{} layer, see SI (Fig.~S2). This system has previously been investigated in detail in Ref.~\citenum{Murray2019}. 

An STM image of \mo{} on \pG{} is shown in Fig.~\ref{fig1}c. O was intercalated after the growth of \mo, as the O would etch Gr at the high temperatures employed in \mo{} growth. The process is detailed in the Methods section. The Gr layer is partially combusted, leaving uncovered areas (holes) in the graphene sheet, while the O atoms intercalate the subsisting Gr. An area where Gr was combusted is partly imaged in the top right of  Fig.~\ref{fig1}c. After the process extended Gr patches remain, on which \mo{} is found. The Gr layer is completely intercalated with a O-($2 \sqrt3 \times 2\sqrt3$)-R$30^{\degree}$ superstructure reported previously, see SI (Fig.~S1)~\cite{Martinez-Galera2016}. Since O-intercalation decouples Gr from Ir(111), the Gr/Ir(111) moir\'e vanishes and a moir\'e between MoS$_2$ and Gr becomes visible as shown in the inset of Fig.~\ref{fig1}c.

		\begin{figure*}
		\centering
		\includegraphics[width=0.55\textwidth]{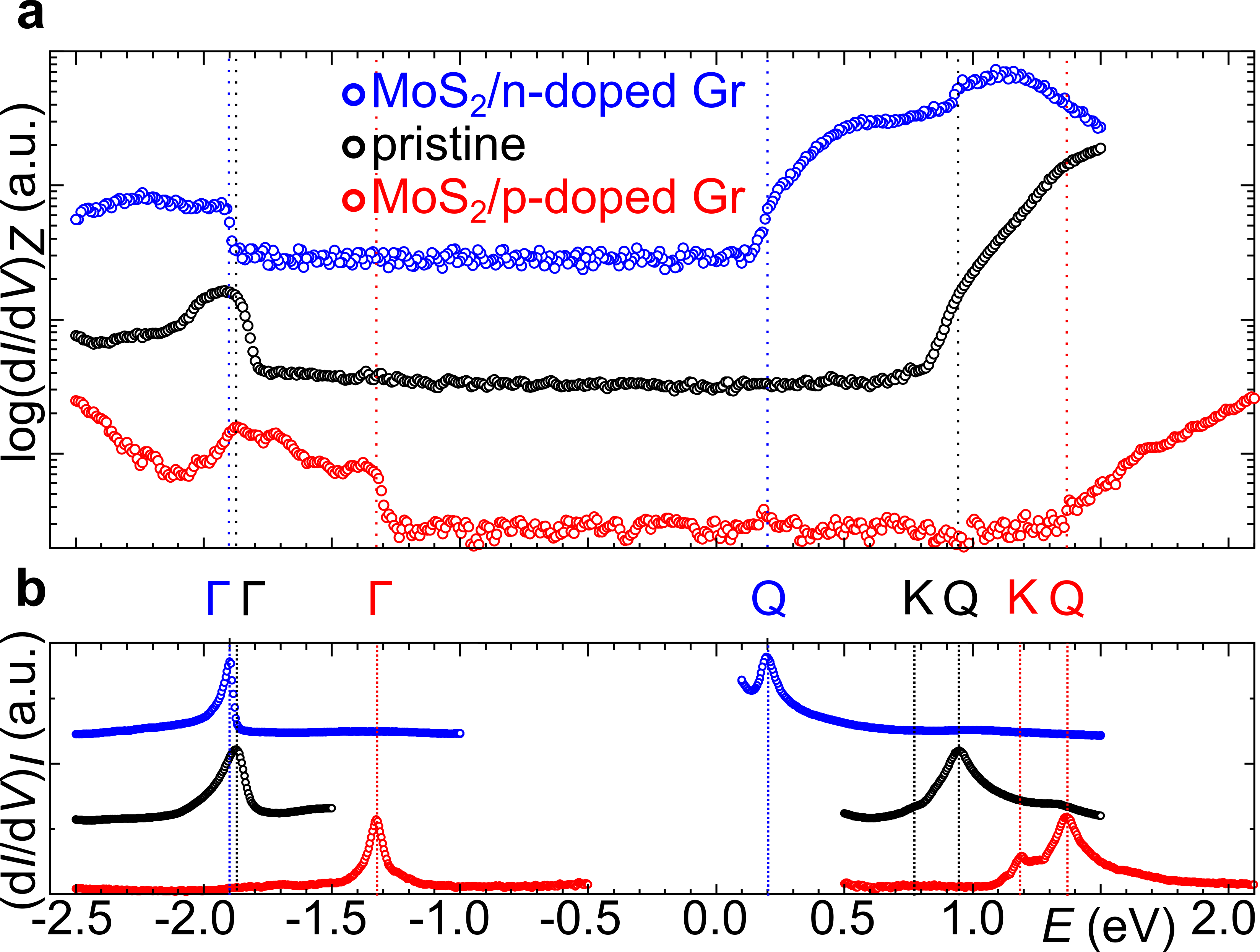}
		\caption{\footnotesize STS of ML \mo{} on doped Gr, measured in  \textbf{a}~constant height and  \textbf{b}~constant current STS modes. \mo{} on Gr/Ir is shown in black; on Gr/O/Ir in red; on Gr/Eu/Ir in blue. The critical point energies are identified. The spectra have been shifted in $\text{d}I/\text{d}V$ for visual clarity. STS parameters:
		 \textbf{a}~$V_\text{st}$= \SI{1.5}{\V}, $I_\text{st}$=\SI{0.2}{\nA} (black); \SI{2.5}{\V}, \SI{0.5}{\nA} (red); \SI{1.5}{\V}, \SI{0.3}{\nA} (blue); 	 
		  \textbf{b}~$I_\text{st}$=\SI{0.1}{\nA} (black); \SI{0.1}{\nA} (red); \SI{0.2}{\nA} (blue).
		 \label{figcompSTS}}
	\end{figure*}

Fig.~\ref{fig1}d is an STM topograph of the surface morphology of \mo{} on \nG. The Gr layer not covered by \mo{} is not uniformly intercalated: large intercalation patches and stripes of apparent height \SI{0.30}{\nm} are visible, similar to previous Gr/Eu/Ir(111) samples \cite{Schumacher2013}. All investigated \mo{} islands have an apparent height of around \SI{0.70}{\nm} relative to the higher [brighter in d] patches of Gr, under similar tunneling conditions as for the pristine sample. We thus infer that ML \mo{} grows preferentially on top of Gr/Eu/Ir(111). A similar preference for adsorption on Eu-intercalated patches was observed for aromatic molecules~\cite{Huttmann2015}. The moir\'e of Gr/Ir(111) is no longer visible in the \mo{} layer; instead, the layer shows slight apparent height modulations, most likely due to the Eu intercalation patterns underneath, see SI (Fig.~S2).

We use STS to investigate how the band structure of \mo{} is altered due to doping of the Gr substrate. The spectra in Fig.~\ref{figcompSTS}a and Fig.~\ref{figcompSTS}b are recorded in constant height and constant current mode on \mo, respectively. Constant height STS, in Fig.~\ref{figcompSTS}a, is proportional to the local density of states and hence records the band gap position in energy. For pristine \mo{} (black dots), the apparent valence band (VB) and conduction band (CB) edges are close to \SI{-1.8}{\eV} and \SI{0.9}{\eV}, respectively. The doping of Gr leads to non-trivial energy shifts of these apparent band edges. The spectrum measured on \mo{} on \pG{} displays a nearly rigid upward shift (\SI{\approx 450}{\milli\eV}) of the apparent VB and CB edges (red dots). Since both band edges shifted by approximately the same amount, the p-doping of Gr leaves the \mo{} band gap nearly unchanged with only a minor reduction of \SI{\approx 20}{\milli\eV}. Upon n-doping, a considerable narrowing of the band gap is found, with the apparent CB edge shifting close to the Fermi energy, while the apparent VB edge remains at approximately the same location (blue dots). The shift of the apparent CB by \SI{\approx 700}{\milli\eV} demonstrates a large shift of the chemical potential of \mo, whereas the lack of a significant shift of the apparent VB is indicative of a large band gap renormalization, as will be analyzed in more detail below.
		
Constant current STS detects the critical point energies in the band structure~\cite{Zhang2015,Murray2019}. It serves as a complementary method to constant height STS. The resulting spectra are shown in Fig.~\ref{figcompSTS}b. We follow the analysis of Ref.~\citenum{Murray2019} to identify the critical point energies of \mo. In the VB, the $\Gamma$-point band edge states cause a large peak \cite{Murray2019}, which is found in all three systems. The actual VB maximum at the K-point is not detected in \mo{} because of its large parallel momentum, in-plane orbital character and close proximity in energy. Based on ARPES measurements, the VB K-point was estimated to lie \SI{0.11}{\eV} above the $\Gamma$-point energy of the VB~\cite{Ehlen2019, Ehlen2017}. The CB minimum at the K-point, however, is dominated by out-of-plane orbitals and thus appears as a small shoulder \cite{Komsa2013_PRB, Murray2019}. This shoulder is found in the pristine sample and in \mo{} on \pG, but cannot be detected in \mo{} on \nG{} with constant current STS, as the signal diverges near the Fermi energy. States from the CB Q-point edge cause a larger peak, which is reproducibly found in all three samples. For that reason, we specify the gap size $E_{\Gamma-\text{Q}}$, used for determining the renormalization energy, as the energetic distance between the well-defined VB $\Gamma$ peak and the CB Q peak in constant current STS. The actual bandgap is expected to be at least \SI{240}{\milli\eV} smaller, see Ref.\,\citenum{Murray2019}. The exact location of the critical point energies and estimated $E_{\Gamma-\text{Q}}$ gap sizes are listed in Table~\ref{tbl:CPEs}.

\begin{table}
		\caption{\footnotesize Critical point energies (eV) identified in ML \mo~on different substrates using constant current STS (averaged over multiple data sets), and the corresponding distance between the well-defined $\Gamma$ and Q peaks in STS $E_{\Gamma-\text{Q}}$. For more details on the \mo/Gr/Ir data, including the determination of the actual  electronic bandgap $E_{\text{g}}=2.53$\,eV, see Ref.\,\citenum{Murray2019}. The location of the K-point of \mo{} on \nG{} has been determined using quasiparticle interference mapping (see Fig.~\ref{figmetal}), since it is not accessible \textit{via} constant current STS.}
		\label{tbl:CPEs}
		\begin{tabular}{c|cccc}
			\mo~on & $\Gamma$ & K & Q & $E_{\Gamma-\text{Q}}$ \\ 
			\hline
			Gr/Ir & $-1.87\pm0.02$  & $0.77\pm0.02$ & $0.90\pm0.05$ & $2.77\pm0.05$ \\ 
			Gr/Eu/Ir & $-1.90\pm0.01$  &$ (-0.09\pm0.01)\hphantom{-}$& $0.20\pm0.01$ & $2.10\pm0.01$ \\ 
			Gr/O/Ir & $-1.37\pm0.04$  & $1.22\pm0.03$ & $1.38\pm0.03$ & $2.75\pm0.05$ \\ 
		\end{tabular}
\end{table}

	\begin{figure*}
		\centering
		\includegraphics[width=0.45\textwidth]{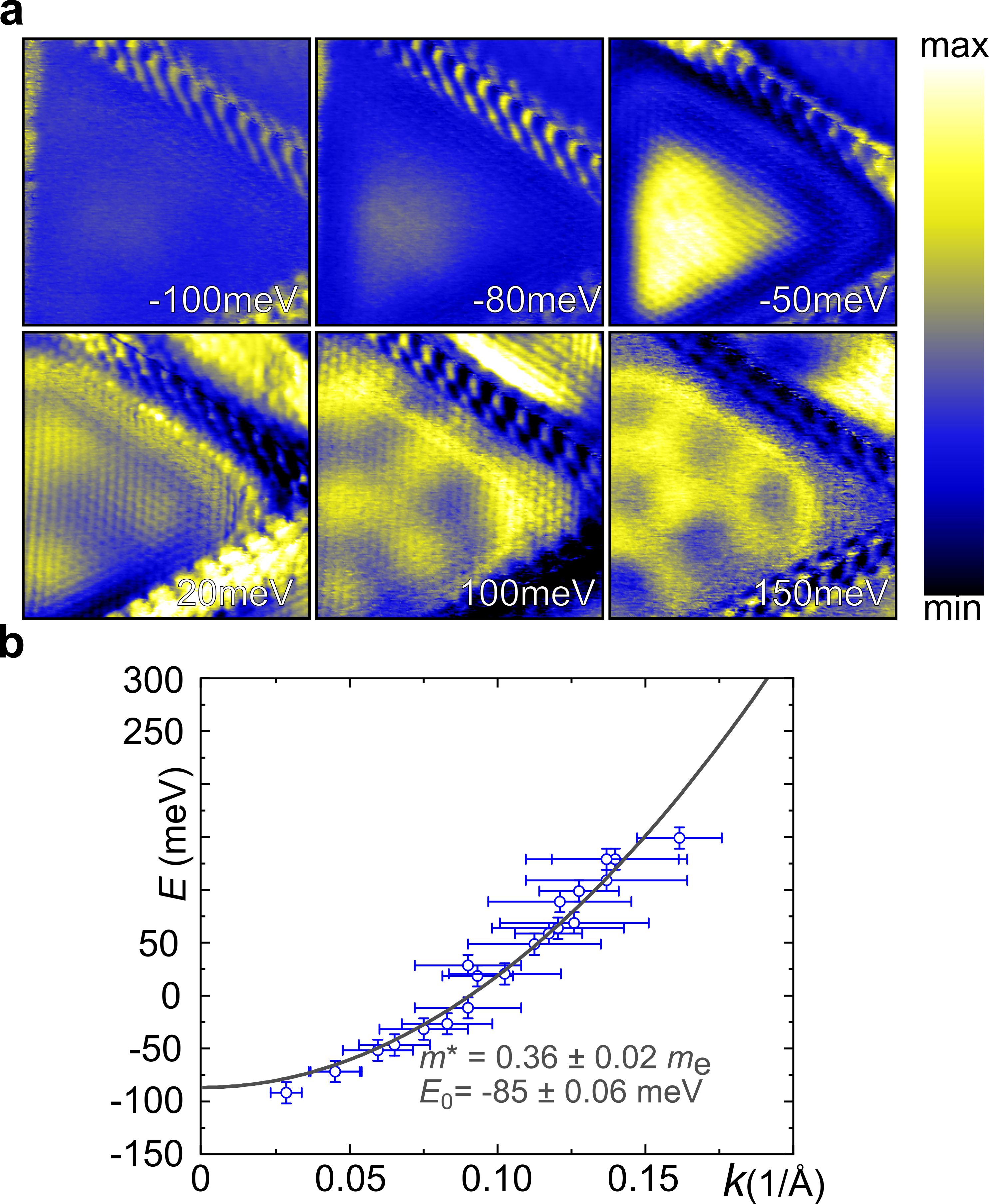}
		\caption{\footnotesize Metallic \mo\, on Gr/Eu/Ir(111). \textbf{a}~Six constant current differential conductance maps in an energy range of \SI{-100}{\milli\eV} to \SI{150}{\milli\eV} with increasing energy. Imaged is a triangular quantum well formed by two MTBs and an island edge enclosing a small area of \mo. \textbf{b}~ Using multiple sets of STS maps, the dispersion of the electron-like conduction band near the K-point is plotted with a best parabolic fit. 
		STM/STS parameters:
		 \textbf{a}~$V_\text{st}$ as indicated, $I_\text{st}$=\SI{0.1}{\nA}, image size \SI{9 \times 9}{\nano\meter\squared}.	 
		 \label{figmetal}}
	\end{figure*}

Remarkably, we find that the CB of \mo{} on \nG{} is effectively at the Fermi level. To address this point, we perform STS mapping to investigate the conduction band minimum (CBM) of \mo{} on \nG. \mo{} patches enclosed by grain boundaries and island edges host confined states, effectively forming a quantum well. Mapping these standing waves allows us to detect the quasiparticles close to the Fermi energy. Fig.~\ref{figmetal}a displays STS maps of such a triangular quantum well, in which we identify quantized states~\cite{Li1985}. The lowest state is pronounced already at \SI{-50}{\milli\eV} below the Fermi energy, providing the smoking gun for the metallic nature of \mo{} on this substrate. In addition, more complex states appear at higher energies, as expected for an electron-like band.
	
Quasiparticle interference patterns around defects and confined quantum well states such as those in Fig.~\ref{figmetal}a are used to determine the dispersion of the quasiparticles around the Fermi energy. The result is shown in Fig.~\ref{figmetal}b. The different data sets and methods used to obtain these data points are shown and discussed in the SI (Fig.~S3). We fit the data with a parabolic dispersion and find that the bottom of the band is located at \SI{-85 \pm 6}{\milli\eV}. This is consistent with constant height STS measurements taken with small tip-sample distances, shown in the SI (Fig.~S4). We further obtain an effective mass of \SI{0.36 \pm 0.02}{\electronmass}, which agrees with the calculated effective mass of \SIrange{0.35}{0.40}{\electronmass} at the K-point~\cite{Cheiwchanchamnangij2012, Qiu2013a}.

%%%%%%%%%%%%%%%%%%%%%%%%%%%%%%%%%%%%%%%%%%%%%%%%%%%%%%%%%%%%%%%%%%%%%%%%%%%%%%%%%%%%%%%%%%%%%%%%%%%%%%%%%%%%%%%%%%%%

\subsection{DFT calulations and model of band shifts}

\begin{figure*}
		\centering
		\includegraphics[width=0.9\textwidth]{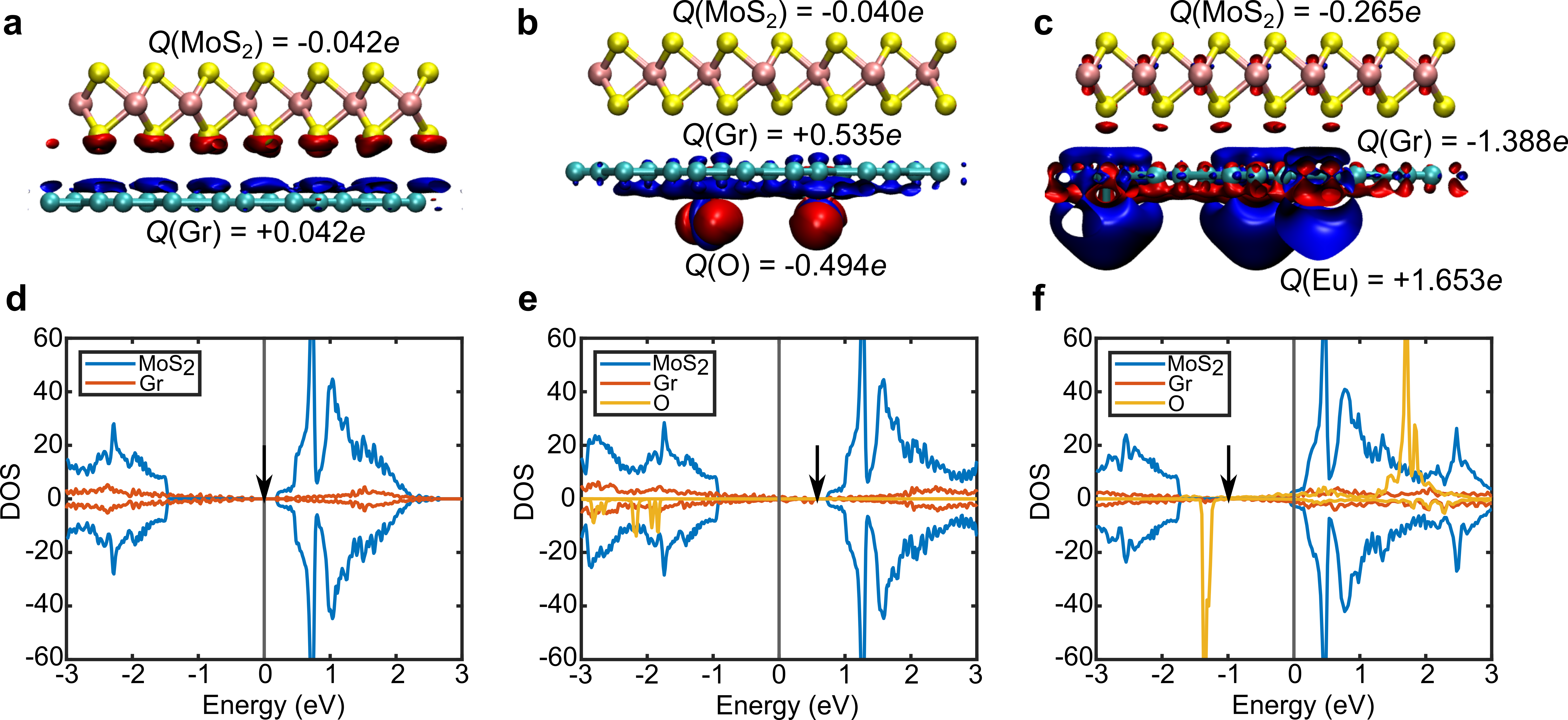}
		\caption{\footnotesize Density functional theory calculations of chemically gated \mo.
                  Atomic structures of \textbf{a}~\mo/Gr, \textbf{b}~\mo/Gr/O and \textbf{c}~\mo/Gr/Eu heterostructures. $Q$ denotes the total charge (sum of Bader charges) in each of the constituent layers. The charge transfer is further visualized by the red and blue isosurfaces for increased and decreased electron density, respectively. The isosurface values are a 0.001 and (b,c) \SI{0.003}{\e/\angstrom^3}. \textbf{d--f}~The corresponding local density of states. Positive and negative values refer to spin-up and -down components, respectively. Vertical black arrow denotes the position of the Gr Dirac-point. Fermi level is set to zero.}
		 \label{figdft}
	\end{figure*}

We conducted DFT calculations to obtain a qualitative understanding of the mechanism through which the shifts of the chemical potential of \mo{} are enabled. Our supercell includes the \mo{} and Gr sheets, while the Ir substrate was omitted. This simplified model of the experimental situation grasps the essentials at affordable computational effort. We also note that the renormalization of the fundamental bandgap due to screening is not accounted for in our DFT calculations. Since the band alignment is sensitive to strain in \mo{} and Gr, we adopted a supercell model where strain was minimized.

Figs.~\ref{figdft}a--c show the structural models of the pristine heterostructure and the intercalated structures containing two O and two Eu dopants, respectively. The figures also show the total charge transfers, as obtained by summing up the Bader charges, and the charge density difference isosurfaces [defined as $\rho({\rm MoS_2/Gr})-\rho({\rm MoS_2})-\rho({\rm Gr})$ and $\rho({\rm MoS_2/Gr/dopants})-\rho({\rm MoS2/Gr})-\rho({\rm dopants})$]. In the case of the pristine heterostructure, the electronic structure at the interface is slightly perturbed by the presence of the other layers, which effectively yields a small charge transfer and an interface dipole induced potential shift of about 0.25 eV.
In the case of the O-doped heterostructure, there is significant electron transfer from the Gr to O, while the \mo{} layer is left mostly unaffected. This is in stark contrast to the Eu-doped heterostructure, where we find charge transfer from Eu to both Gr and \mo, albeit with the charge transfer to Gr still considerably larger. The charge transfer, induced dipoles and Dirac-point positions are listed in the SI (Table~S1).

Figs.~\ref{figdft}d--f show the DFT-calculated local densities of states projected onto the two or three constituent layers. We find that pristine \mo~is semiconducting, with the Fermi level 0.2 eV below CBM and the gap of 1.69 eV close to the value calculated for an  isolated monolayer~\cite{Komsa2015}. Gr remains charge neutral, with the Fermi level coinciding with the Dirac-point, since the interface dipole is the result of modifications to the extensions of the Gr wave functions and not of charge transfer. 
In the case of the O-doped heterostructure Gr is p-type doped, with the Fermi level 0.58 eV below the Dirac-point and 0.75 eV below the \mo~CBM. Thus we find a near-rigid shift of the Gr and \mo{} bands to higher energies, in agreement to our experiments. In the case of the Eu-doped heterostructure Gr is n-type doped, with the Fermi level 0.99 eV above the Dirac-point and 0.05 eV above \mo~CBM. This represents a strong deviation from a rigid shift picture, in which one would expect the Fermi level to lie 0.8~eV above the \mo~CBM. DFT thus reproduces the metal--insulator transition, as well as the shifts of the Gr Dirac point and \mo{} conduction band found experimentally, including deviations from a rigid shift. Our DFT model thus captures the essence of our experimental observations, although the absence of the Ir substrate leads to a significantly lower density of dopants needed to obtain shifts comparable to the experiments.

	\begin{figure*}
		\centering
		\includegraphics[width=0.45\textwidth]{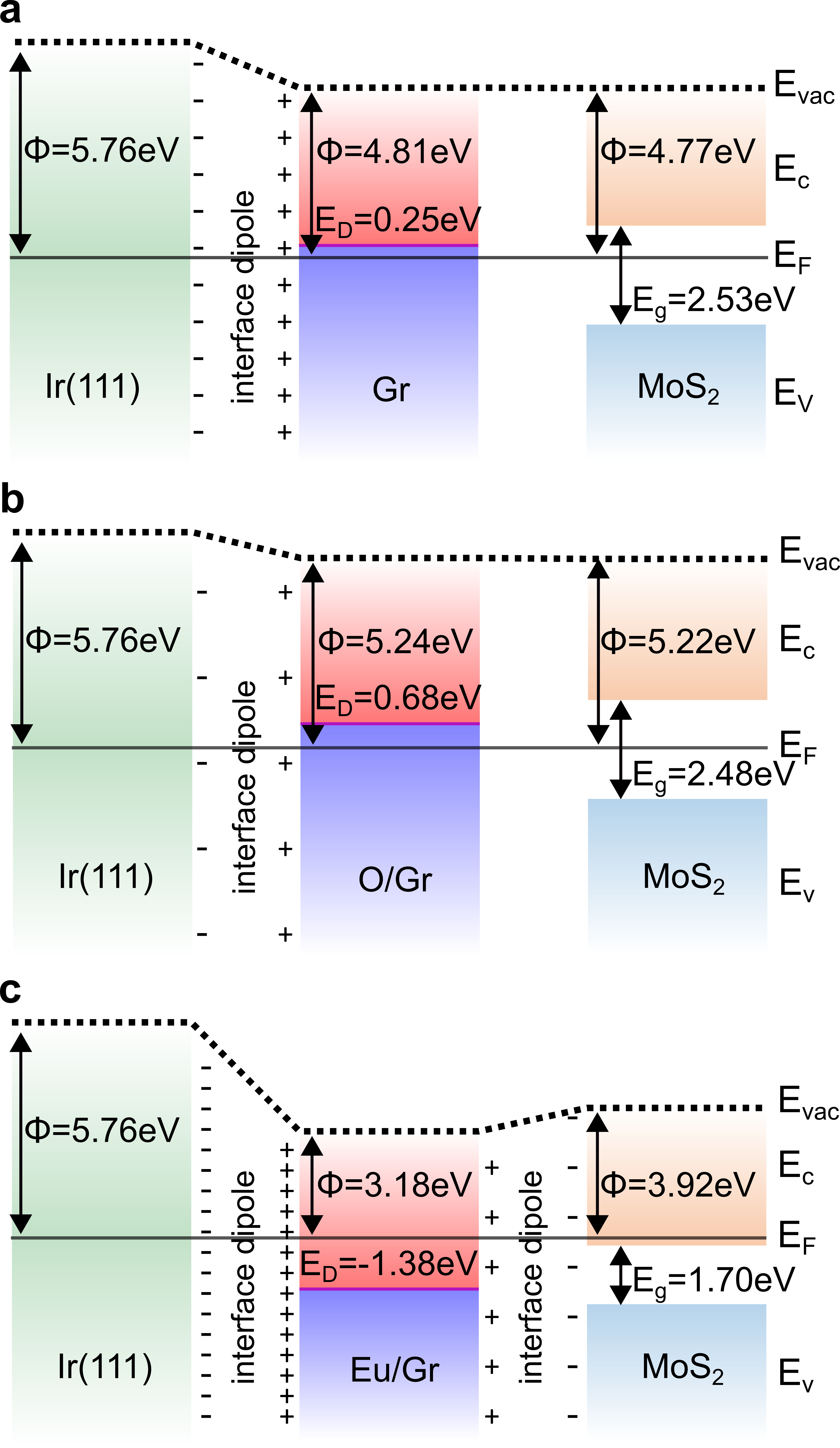}
		\caption{\footnotesize Band diagrams of \mo{} on differently charged substrates. \textbf{a}~\mo/Gr/Ir(111). \textbf{b}~\mo/Gr/O/Ir(111). \textbf{c}~\mo/Gr/Eu/Ir(111). Indicated are the vacuum level $E_{\text{vac}}$, the Fermi level $E_{\text{F}}$, the workfunctions $\Phi$, the Gr Dirac point $E_{\text{D}}$ and the \mo{} conduction and valence bands $E_{\text{C}}$ and $E_{\text{V}}$, respectively. 
			\label{figdip}}
	\end{figure*}

The physics involved in the changes of band edge positions and bandgap width can be grasped in a model, accounting for workfunction alignment and interface dipole formation, as shown in Figs.~\ref{figdip}a--c. The \mo{}, Gr and Ir(111) band diagrams are sketched for the three samples. Two interfaces are considered: between (doped) Gr and Ir(111), and between (doped) Gr and \mo. If charge transfer is possible between the layers (\textit{e.g.} the metal--metal junction between Gr and Ir), an interface dipole will be formed, as a result of workfunction differences. If no charge transfer is possible because one of the materials is gapped (\textit{e.g.} the metal--semiconductor junction between undoped Gr and \mo), the workfunctions of the two materials have to align through shifts in the position of the chemical potential with respect to the band structure and thus to the vacuum level. 

Both types of interfaces are present in the pristine case sketched in Fig.~\ref{figdip}a. The Ir workfunction $\Phi_{\text{Ir}} = \SI{5.76}{\eV}$~\cite{Michaelson1977} is considerably larger than the workfunction of pristine Gr $\Phi_{\text{Gr}} = \SI{4.56}{\eV}$~\cite{Yan2012}. Due to the difference in workfunctions, charge will flow from Gr to Ir until an interface dipole is formed, halting the flow of further charge carriers. As a result, Gr is slightly p-doped when grown on Ir(111), by about \SI{0.1}{\eV}~\cite{Pletikosic2009}. Note that for physisorbed Gr with an intact Dirac cone, \textit{e.g.} Gr on Ir(111), even strong doping does not change the energetic separation of the Dirac point from the vacuum level, it being identical to the workfunction of undoped Gr~\cite{Khomyakov2009}.

When ML \mo{} is placed on top of Gr, the interface dipole between Gr and Ir is marginally changed due to a slight increase in the p-doping of Gr, from \SI{+0.1}{\eV} to \SI{+0.25}{\eV}~\cite{Ehlen2019}, resulting in a Gr workfunction $\SI{4.56}{\eV} + \SI{0.25}{\eV} = \SI{4.81}{\eV}$. Between the interior of the \mo{} layer and Gr no charge transfer is possible, as the Fermi level lies within the \mo{} band gap. Therefore, with no dipole formed, the workfunctions of both materials have to align. The chemical potential of \mo{} correspondingly shifts from the middle of the band gap closer to the CB. With the ML \mo{} electron affinity $\chi \SI{\approx4.0}{\eV}$~\cite{Das2013, Lee2016d, Hu2018, Zhang2016a, Schlaf1999}, and finding the conduction band edge $E_{\text{c}}$ of \mo/Gr/Ir(111) in STS at \SI{0.77}{\eV}, the workfunction of \mo{} on Gr/Ir(111) $\Phi_{\text{\mo}}=\chi+E_{\text{CBM}}=\SI{4.77}{\eV}$ is indeed very close to that of Gr: $\Phi_{\text{Gr}} = \SI{4.81}{\eV}$.
 
When O is intercalated underneath Gr, shown in Fig.~\ref{figdip}b, the Dirac point moves to \SI{0.68}{\eV}~\cite{Jolie2014} and the Gr workfunction changes to $\SI{4.56}{\eV} + \SI{0.68}{\eV} = \SI{5.24}{\eV}$. The interface dipole between Gr and Ir becomes smaller. Following the previous considerations, the \mo{} bands must follow the shift of the Dirac point, in order to keep the workfunctions of Gr and \mo{} the same. The measured K-point shift in the \mo{} CB of $\SI{\approx 450}{\milli\eV}$ is correspondingly close to the expected Gr Dirac point shift of $\SI{430}{\milli\eV}$. The chemical potential of \mo{} is thus directly set by that of Gr, as long as \mo{} remains insulating. 

Experimentally, we find that the size of the band gap $E_{\text{g}}$ is only slightly reduced, when comparing the p-doped and pristine sample. Since \mo{} is still insulating, the screening within the layer should not have changed. Gr, on the other hand, has an increased hole density on the order of \SI{10^{13}}{\cm^{-2}} due to an estimated Dirac point energy $E_{\text{D}}$ = \SI{-0.68}{\eV}~\cite{Larciprete2012}. Gr is thus expected to better screen electrostatic interactions in \mo. Previous experimental and theoretical work on TMDC/Gr heterostructures has however found that the effect of increased Gr carrier density on the band gap flattens off after $E_{\text{D}}$ =\SI{\pm 0.25}{\eV}~\cite{Qiu2019, Riis-Jensen2020}. Since the Gr layer is already p-doped in the pristine \mo{} sample~\cite{Ehlen2019}, the observed renormalization of $\Delta E_{\text{g}} \approx \SI{20}{\milli\eV}$ seems to indicate that the substrate-induced renormalization is already close to saturation prior to O-intercalation. 

For Eu-doped Gr, Fig.~\ref{figdip}c, the situation is decidedly different. Gr/Eu has its Dirac point at about \SI{-1.38}{\eV}, see Ref.~\citenum{Schumacher2013a} and the SI (Fig.~S5). The chemical potential difference between Eu-doped Gr and pristine \mo{} is larger than the difference between pristine \mo{} and the pristine CB edge. The \mo{} potential will therefore shift in response to the Eu-doped Gr potential until it hits the CB and \mo{} becomes metallic. Charge flows from Gr to \mo{}, creating an interface dipole, which reduces the shift of the \mo{} CB minimum compared to the shift of the Dirac point of graphene.

In addition to the formation of an interface dipole, the charge carriers within \mo{} are also able to efficiently screen within the layer, leading to the experimentally observed band gap reduction $\Delta E_g = \SI{0.67}{\eV}$. The renormalization of the bandgap due to the metal--insulator transition is considerably larger than what is achieved \textit{via} methods which only increase the carrier density in the substrate. In those cases, renormalization energies of at most \SI{0.24}{\eV} have been reported~\cite{Ugeda2014, Qiu2019}.

%%%%%%%%%%%%%%%%%%%%%%%%%%%%%%%%%%%%%%%%%%%%%%%%%%%%%%%%%%%%%%%%%%%%%%%%%%%%%%%%%%%%%%%%%%%%%%%%%%%%%%%%%%%%%%%%%%%%

\subsection{Effect of charge on mirror twin boundaries}\label{sec1Din}
The presence of 0D defects (such as vacancies) or 1D defects (such as MTBs) in TMDCs will influence the shifts of the chemical potential induced via contactless chemical doping. In-gap states within these defects tend to weaken the impact of the work function difference due to local Fermi level pinning~\cite{Allain2015, Diaz2016, Bampoulis2017}, which is absent away from these defects~\cite{LeQuang2017}. In the pristine sample, the presence of in-gap states below the Gr Fermi level in \mo{} edges and MTBs leads to the accumulation of charge in the MTBs, causing band bending in the surrounding semiconductor~\cite{Murray2020}. With the doping of Gr potentially influencing the amount of charge that is transfered to the MTBs as well as the dielectric properties of the \mo/Gr heterostructure (\textit{e.g.} by inducing a metal--insulator transition), it is worthwhile to disentangle the various effects of Gr doping on the MTB states and the electronic landscape surrounding them. 

	\begin{figure*}
		\centering
		\includegraphics[width=0.9\textwidth]{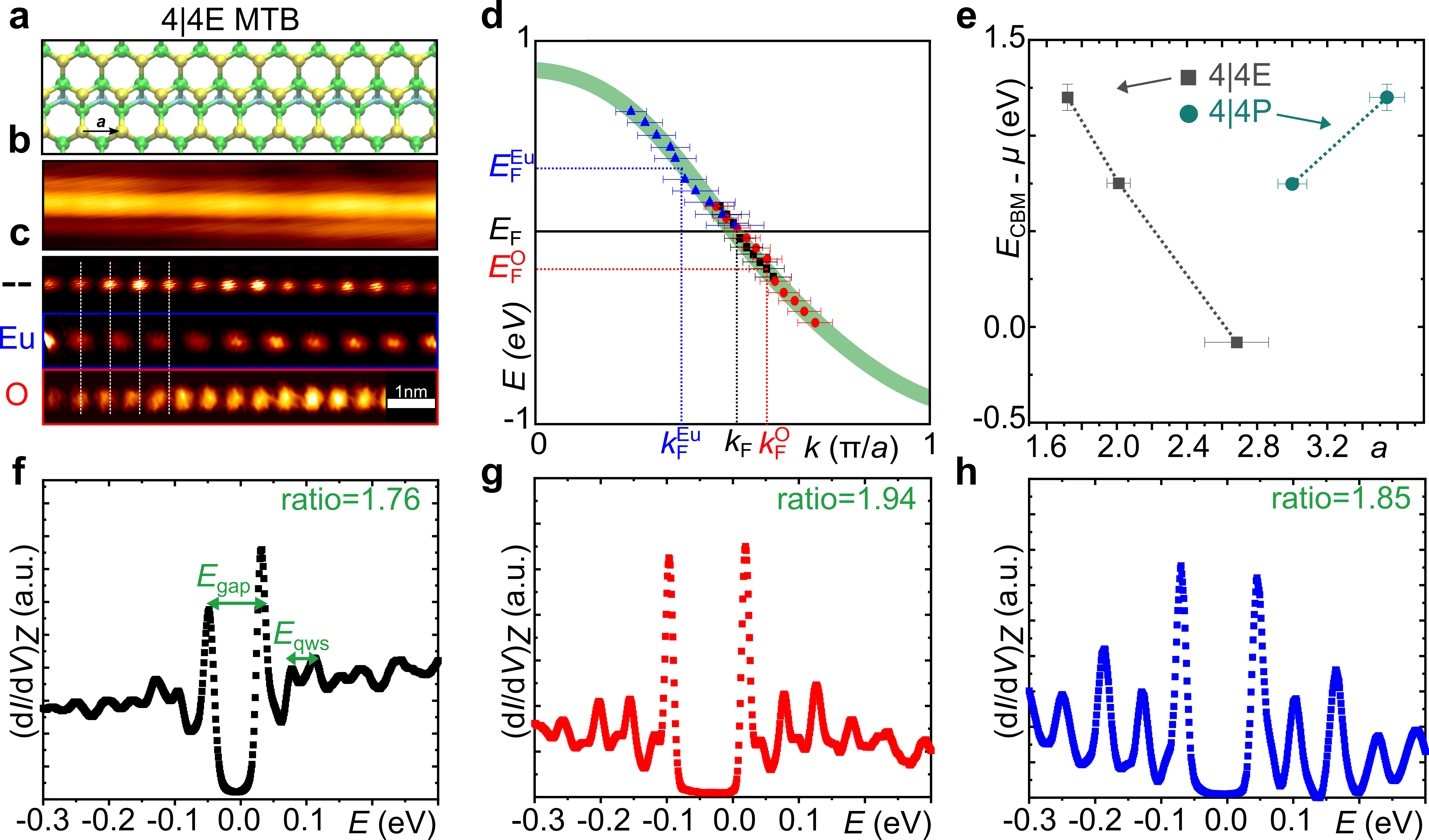}
		\caption{\footnotesize Confined states along 4|4E and 4|4P MTBs in ML \mo{} on different substrates. 
		\textbf{a}~Ball-and-stick model of a 4|4E MTB. Mo atoms are navy, S atoms yellow (top layer) or dark yellow (bottom layer). A \mo{} lattice vector is indicated.
		\textbf{b}~Constant current STM topograph of a 4|4E MTB in \mo{} on Gr/Ir(111). 
		\textbf{c}~Constant height STM image of a 4|4E MTB in pristine (--), p-shifted (O) and n-shifted (Eu) \mo. Vertical white lines are a guide to the eye. 
		\textbf{d}~Dispersion of the hole-like 1D band of the 4|4E MTB, with data from MTBs on the differently doped Gr substrates in their respective colors. In green a DFT-calculated dispersion for \mo/Gr/Ir(111) is shown, taken from Ref.~\citenum{Murray2020}. Experimental data from spectra taken along the same MTBs as used for the averaged linescans in \textbf{f--h}. 
		\textbf{e}~Plot of 4|4E (gray) and 4|4P (teal) MTB beating periodicity near $E_\text{F}$ versus the location of the CBM. The dotted lines are a guide to the eye. 
		\textbf{f--h}~Constant height STS linescans along a 4|4E MTB in \mo/Gr/Ir(111) (left), \mo/Gr/O/Ir(111) (middle) and \mo/Gr/Eu/Ir(111) (right). The spectra are averaged over the whole length of the MTB. Indicated is the ratio of the gap at the Fermi level ($E_{\text{gap}}$) with respect to the average gap between the quantized states ($E_{\text{qws}}$).
		STM/STS parameters: \textbf{a}~top panel: $V$=\SI{1.25}{\V}, $I$=\SI{0.02}{\nA} (left); --:$V$=\SI{0.10}{\V}, $I_\text{st}$=\SI{0.5}{\nA}; O: $V$=\SI{0.02}{\V}, $I_\text{st}$=\SI{0.1}{\nA}; Eu: $V$=\SI{0.20}{\V}, $I_\text{st}$=\SI{0.05}{\nA}; \SI{1}{\nm} scale bar shown. \textbf{f}~$V_\text{st}$=\SI{0.5}{\V}, $I_\text{st}$=\SI{0.1}{\nA}; \textbf{g}~$V_\text{st}$=\SI{0.5}{\V}, $I_\text{st}$=\SI{0.1}{\nA}; \textbf{h}~$V_\text{st}$=\SI{-0.5}{\V}, $I_\text{st}$=\SI{0.1}{\nA}.
			\label{figqs}}
	\end{figure*}
	
In \mo{} two types of MTBs (4|4E and 4|4P) are present, formed between $180^\circ$-misoriented \mo~domains \cite{Zou2013,Komsa2017}. A model and a constant currrent STM image of the 4|4E MTB are shown in Fig.~\ref{figqs}a and Fig.~\ref{figqs}b, respectively. The 4|4P MTB is discussed in the SI (Fig.~S4). In the pristine sample, the states within MTBs are strictly 1D, leading to strong electron-correlations and the emergence of Tomonaga-Luttinger liquid (TLL) behaviour~\cite{Jolie2019a}. In addition, since the states are located in boundaries of finite length, they are confined. Hence, when mapping the local density of states close to the Fermi energy, we expect a pronounced, sinusoidal beating pattern along the length of the MTB, resulting from quantized 1D electronic states, as shown in Fig.~\ref{figqs}c. The wavelength of this beating pattern is related to the confined states closest to the Fermi energy, $\lambda=\pi/k_\text{F}$~\cite{Jolie2019a}. Therefore, if the charge donated by the Gr substrate shifts the Fermi level of the MTB band, the change in filling can be observed directly with STM, by imaging the periodicity of the states near $E_{\text{F}}$.

The constant height STM images in Fig.~\ref{figqs}c, taken near $E_{\text{F}}$, reveal the beating patterns found within 4|4E MTBs for \mo{} on the differently doped substrates. In the pristine case, we find $\lambda_\text{F}^{\text{Gr}}=(2.01\pm0.04)\,a$ for the 4|4E MTBs, where $a$=\SI{0.315}{\nm} is the ML \mo{} lattice constant~\cite{Kormanyos2015}. For p-shifted \mo{} the periodicity along the 4|4E MTBs has visibly decreased, $\lambda_\text{F}^{\text{O}}=(1.72\pm0.01)\,a$, while for n-shifted \mo{}, an increased periodicity of $\lambda_\text{F}^{\text{Eu}}=(2.69\pm0.11)\,a$ is measured.

The shifts in the beating periodicity can be understood from the band structure of the MTB. 4|4E MTBs host a hole-like band, which is shown in green in Fig.~\ref{figqs}d, based on DFT calculations discussed in Ref.~\cite{Jolie2019a}. With the Fermi level $E_{\text{F}}$ in the pristine case bisecting the band at $k_{\text{F}}=\pi/2a$, this leads to the observed $2a$ periodicity. When the Gr substrate is n-doped, additional charge can flow into the boundary, raising the Fermi level to $E_{\text{F}}^{\text{Eu}}$ and leading to an increased wavelength $\lambda_\text{F}^{\text{Eu}}$. For p-doping of Gr, the inverse happens: charge is extracted from the boundary, the Fermi level sinks to $E_{\text{F}}^{\text{O}}$ and $\lambda_\text{F}^{\text{O}}$~decreases. The quantitative relationship between $k_{\text{F}}$ and the line charge in the boundary is given in the SI (Table~S2) . 

From the measured $k$ values we find that the chemical potential shifts in the MTB are considerably smaller than those in Gr or \mo, with the Fermi level difference between the 4|4E on n- and p-doped Gr about \SI{500}{\milli\eV}, see SI (Table~S2) for the exact values. In other words, the states within MTBs locally pin the Fermi level, while Fermi level pinning is absent in \mo{} away from defects. This can be understood from previous considerations: the metal-metal interface between the MTB and Gr allows for charge transfer and thus the formation of an interface dipole. As a consequence, the MTB Fermi level will not shift as much as the chemical potential of \mo, which rigidly follows the Gr doping level until it becomes metallic. Note that the relation between $\lambda_{\text{F}}$ and Fermi level is reversed for 4|4P MTBs, which host an electron-like band, see SI (Fig.~S6).

Our findings are summarized in Fig.~\ref{figqs}e in which we plot the position of the CBM of \mo{} as a function of beating period for the two types of MTBs. This allows the identification of the Fermi wave vector of the MTB band and the position of the chemical potential $\mu$ with respect to the \mo{} conduction band in a single STM image. Due to the close proximity of CBM and chemical potential in our n-shifted \mo{} samples, we can also set a lower bound of $\lambda_\text{F}^{\text{Eu}}=(2.69\pm0.11)\,a$ for the beating pattern within 4|4E MTBs, for which the surrounding 2D material is metallic. As a consequence, the MTBs function as a sensor - they allow for a determination of the chemical potential of \mo{} by measuring the periodicity of the MTB states near $E_\text{F}$. This method is also applicable at room temperature.

In Fig.~\ref{figqs}f--h we display additional constant height STS linescans averaged along the length of 4|4E MTBs on all three substrates, respectively. The finite length of the MTBs leads to a series of peaks in the d$I/$d$V$ spectra at the energetic location of the quantized states. The dominant lowest energy excitations for a given wavevector are used to obtain the dispersion within the 4|4E MTBs, see Fig.~\ref{figqs}d~\cite{Jolie2019a}. We find a pronounced gap $E_{\text{gap}}$ around the Fermi level, consistent with the presence of a Coulomb blockade which pushes the states closest to $E_\text{F}$ further apart~\cite{Jolie2019a}. The gap is almost twice as large as the energy spacing between the quantized states $E_{\text{qws}}$. We find that the ratio $E_{\text{gap}}/E_{\text{qws}}$, which eliminates the length-dependence of both parameters~\cite{Yang2021a}, has only minor substrate-dependence. This is surprising, since the presence of a Coulomb gap, which arises from strong electron--electron interactions, indicates that the 1D correlated states within 4|4E MTBs persist even when the surrounding \mo~is metallic.

	\begin{figure*}
		\centering
		\includegraphics[width=0.9\textwidth]{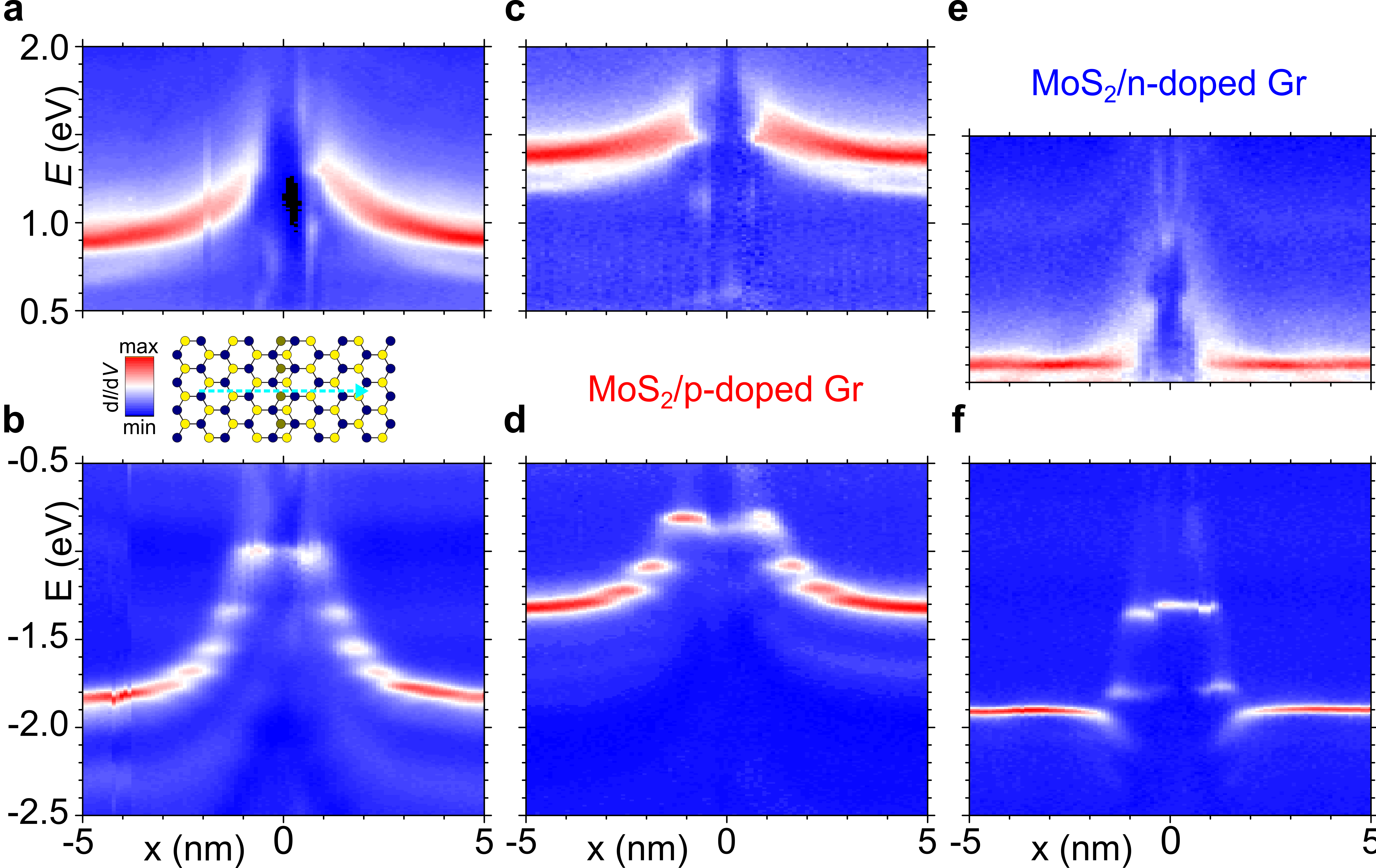}
		\parbox{\textwidth}{\caption{\footnotesize Band bending at 4|4E MTBs in ML \mo{} on doped Gr: \textbf{a-b} on Gr/Ir; \textbf{c-d} on Gr/Eu/Ir; \textbf{e-f} on Gr/O/Ir. Constant current STS linescans are taken in the energy ranges of the CB \textbf{a,c,e} and VB \textbf{b,d,f} extrema. $(\text{d}I/\text{d}V)_I$ signals are plotted as a function of energy $E$ and position $x$ according to the color scale shown in \textbf{a}, in arbitrary units. In a the path of a linescan over a 4|4E MTB model is shown, not to scale. 
		STS parameters:
		\textbf{a,b} $V_\text{st}=\pm2.5$\,V, $I_\text{st}=0.05$\,nA;
		\textbf{c,d} $V_\text{st}=\pm2.5$\,V; 
		\textbf{e} $V_\text{st}=1.5$\,V; d $V_\text{st}=2.0$\,V;
		\textbf{c-f} $I_\text{st}=0.1$\,nA.		
}
			\label{figbb}}
	\end{figure*}

Moving away from the effects of the intercalation on the intrinsic MTB properties, we also investigated the region directly surrounding the boundaries, where the \mo{} bands are bent under the influence of the MTB charge~\cite{Murray2020}. The magnitude of band bending must change if the amount of charge on the MTB is changed in consequence of Gr doping, while the shape of the bending provides insight into how the screening properties of \mo{} are affected.

For reference, we take constant current STS linescans orthogonally crossing a 4|4E MTB in the pristine heterostructure. While Fig.~\ref{figbb}a captures the behaviour of the CB edge,  Fig.~\ref{figbb}b captures that of the VB. Far away from the MTB, at $x=\SI{\pm 5}{\nano\meter}$, the unperturbed pristine band structure is observed. Approaching the MTB, located at $x=\SI{0}{\nano\meter}$, both CB and VB are seen to bend upwards by at least \SI{0.5}{\eV}, indicating that the MTB is negatively charged. The bending is smooth in the CB but occurs stepwise at characteristic energies in the VB due to quantization effects~\cite{Murray2020}. 

On p-doped Gr, Figs.~\ref{figbb}c,d, the magnitude of the bending is severely reduced for both bands, while the spatial extent of the bending is similar to that of the pristine case. There are fewer quantized VB levels and the unperturbed band structure is recovered at $x=\SI{\pm 4}{\nano\meter}$. This can be understood from previous considerations: Gr is more p-doped, lowering its chemical potential. As a result, certain MTB states are now shifted above the Gr Fermi level and less charge can flow from Gr into the MTB, reflected in its smaller periodicity at $E_{\text{F}}$. 

On n-doped Gr, shown in Figs.~\ref{figbb}e, f, the bending of the \mo{} bands occurs in a much smaller range of \SI{|2|}{\nano\meter} on either side of the MTB. In Fig.~\ref{figbb}f there are fewer quantized VB levels -- only two -- and their energy spacing is larger. The highest quantized state occurs at around \SI{-1.3}{\eV}, compared to \SI{-1.0}{\eV} found in the pristine sample. Additionally, the VB bends up steeply at $x \approx \pm 1.5$\,nm from the boundary. The increase in band bending due to graphene n-doping is expected, as more charge is transferred into the MTB states. 

The decrease of the spatial extent of band bending on the one hand is a sign of the metallic phase of \mo{} on Gr/Eu/Ir(111), which effectively screens the charge within the MTB. On the other hand, the steep band bending close to the MTB leaves the direct surrounding of the latter insulating, in line with the persistent Coulomb gap $E_{\rm{gap}}$~found experimentally. 

\section*{Discussion}

We have demonstrated a non-invasive method to strongly modify the electrostatic environment of ML \mo, applicable also to other TMDCs and van der Waals materials. We have shown that the doping of its Gr substrate can induce a metal-insulator transition in \mo{} and enables the manipulation of metallic states within MTBs. In addition, the chemical potential shifts in \mo{} can be monitored \textit{via} MTB states and the screening environment around the boundaries. The backside functionalization of Gr leaves the TMDC top layer chemically pristine, thereby offering advantages over chemical doping or adatom techniques. Our contactless chemical doping method to shift the chemical potential of wide-bandgap semiconductors can be extended by other intercalants such as alkalis~\cite{Petrovic2013}, other rare earth metals~\cite{Klimovskikh2019} or the p-dopant chlorine~\cite{Vinogradov2012}. Continuous tuning of the chemical potential, analogous to gating, could be accomplished by using Li as intercalant~\cite{Schroeder2016, Halle2016}. A combination of contactless gating for coarse adjustment of the chemical potential with electrostatic gating for fine adjustment represents another exciting perspective. We envision that contactless gating will enable observation and characterization of novel states of matter using surface science techniques.

\section*{Methods}
	
	The samples were grown \textit{in situ} in a preparation chamber with base pressure $p < $ \SI{5 \times 10^{-10}}{\milli\bar}. Ir(111) is cleaned by \SI{1.5}{\keV} Ar$^+$ ion erosion and annealing to temperatures $T$ \SI{\approx1550}{\K}. Gr is grown on Ir(111) by two steps. First, room temperature ethylene exposure till saturation followed by \SI{1370}{\K} thermal decomposition gives well-oriented Gr islands. Second, exposure to \SI{2000}{\L} ethylene at \SI{1370}{\K} for \SI{600}{\second} yields a complete single-crystal Gr layer~\cite{Coraux2009}.
	ML \mo{} is grown by Mo deposition in an elemental S pressure of \SI{1 \times 10^{-8}}{\milli\bar}~\cite{Hall2017}.  Subsequently, the sample is annealed to \SI{1050}{\K} in the same S background pressure. 
	The Eu-intercalated sample is prepared in the order: Gr growth; Eu intercalation; \mo{} growth. Eu is evaporated from a Knudsen cell onto the Gr/Ir(111) crystal kept at \SI{720}{\K}. Intercalation is confirmed by low-energy electron diffraction (LEED)~\cite{Schumacher2014}. The sample is flash annealed to \SI{1075}{\K} to remove residual Eu before subsequent \mo{} growth.
	The O-intercalated sample is prepared in the order: Gr growth; \mo{} growth, O intercalation. O intercalation is achieved by exposing \mo/Gr/Ir(111) at \SI{770}{\K} to \SI{2 \times 10^4}{\L} of O$_2$~\cite{Granas2012,Martinez-Galera2016}.
	
	STM and STS are carried out in the $T=$ \SI{7}{\K} bath cryostat system after \textit{in situ} transfer from the preparation chamber. STS is performed with the lock-in technique, at modulation frequency \SI{619-777}{\Hz} and modulation voltage $V_{\text{mod}}=$ \SI{4}{\mV}$_{\text{rms}}$, with an experimental resolution of \SI{\approx10}{\meV} or better \cite{Morgenstern2003}. We employ constant height [recording $(\text{d}I/\text{d}V)_Z$] and constant current [$(\text{d}I/\text{d}V)_I$] STS modes, where $I$ is the tunneling current, $V$ the bias voltage, and $Z$ the tip-sample distance or height. In both modes $\text{d}I/\text{d}V$ is recorded while $V$ is ramped; see Ref.\,\citenum{Murray2019} for further explanation. Spectra are always taken away (\SI{>6}{\nm}) from defects, unless indicated otherwise.
	
	All density functional theory calculations were carried out using the Vienna Ab Initio Simulation Package (VASP) \cite{vasp1,vasp2}. The plane wave cutoff was set to 400 eV throughout. We used the exchange-correlation functional of Perdew, Burke, and Ernzerhof (PBE) \cite{PBE}. Van der Waals interactions are described using Grimme's semi-empirical corrections at the D2 level \cite{GrimmeD2}. A 4$\times$4 k-point mesh was used during structural optimization in the supercell, while the density of states is evaluated using a 16$\times$16 mesh.
        The optimized lattice constants of MoS$_2$ and Gr are 3.18 {\AA} and 2.468 {\AA}, respectively. For a minimally strained heterostructure model, we used the same approach and model as in Ref.~\cite{Komsa2013_PRB}, wherein a 4$\times$4 supercell of MoS$_2$ is interfaced with a (6,3)$\times$(6,3) Gr cell (i.e., $\overline{a}_{\rm sc} = 6\overline{a}+3\overline{b}$, where $\overline{a}$ and $\overline{b}$ are the lattice vectors of Gr at an 120$^\circ$ angle). The dopants were placed as far from each other as possible within the supercell. The optimized geometry for O atoms was between graphene top- and hollow-sites. The optimized geometry for Eu atoms was on the graphene hollow site.

        The MoS$_2$ layer is kept unstrained and Gr layer is compressively strained by 0.8\%. Finally, in order to avoid the buckling of Gr due to chemical interactions with the dopants, charging, or strain, the $z$-coordinates of C atoms were fixed.

\section{Acknowledgments}
This work was funded by the Deutsche Forschungsgemeinschaft (DFG, German Research Foundation) - Project number 277146847 - CRC 1238 (subprojects A01 and B06). Support from the German Academic Exchange Service DAAD and the Academy of Finland \textit{via} PPP Finland MODEST, project IDs 57458732 (DAAD) and 321914 (Academy of Finland) is gratefully acknowledged. The authors thank CSC IT Center for Science for the generous grants of computer time.

\clearpage

	%\bibstyle{achemso}
	\bibliographystyle{naturemag}
	\bibliography{./library}

\begin{thebibliography}{10}
\expandafter\ifx\csname url\endcsname\relax
  \def\url#1{\texttt{#1}}\fi
\expandafter\ifx\csname urlprefix\endcsname\relax\def\urlprefix{URL }\fi
\providecommand{\bibinfo}[2]{#2}
\providecommand{\eprint}[2][]{\url{#2}}

\bibitem{Xi2015}
\bibinfo{author}{Xi, X.} \emph{et~al.}
\newblock \bibinfo{title}{{Strongly enhanced charge-density-wave order in
  monolayer NbSe$_2$}}.
\newblock \emph{\bibinfo{journal}{Nat. Nanotechnol.}}
  \textbf{\bibinfo{volume}{10}}, \bibinfo{pages}{765--769}
  (\bibinfo{year}{2015}).
\newblock
  \urlprefix\url{http://www.nature.com/doifinder/10.1038/nnano.2015.143}.

\bibitem{Chen2020a}
\bibinfo{author}{Chen, Y.} \emph{et~al.}
\newblock \bibinfo{title}{{Strong correlations and orbital texture in
  single-layer 1T-TaSe2}}.
\newblock \emph{\bibinfo{journal}{Nat. Phys.}} \textbf{\bibinfo{volume}{16}},
  \bibinfo{pages}{218--224} (\bibinfo{year}{2020}).
\newblock \urlprefix\url{http://dx.doi.org/10.1038/s41567-019-0744-9}.

\bibitem{Ugeda2016}
\bibinfo{author}{Ugeda, M.~M.} \emph{et~al.}
\newblock \bibinfo{title}{{Characterization of collective ground states in
  single-layer NbSe$_2$}}.
\newblock \emph{\bibinfo{journal}{Nat. Phys.}} \textbf{\bibinfo{volume}{12}},
  \bibinfo{pages}{92--97} (\bibinfo{year}{2016}).

\bibitem{Li2016c}
\bibinfo{author}{Li, L.} \emph{et~al.}
\newblock \bibinfo{title}{Controlling many-body states by the electric-field
  effect in a two-dimensional material}.
\newblock \emph{\bibinfo{journal}{Nature}} \textbf{\bibinfo{volume}{529}},
  \bibinfo{pages}{185--189} (\bibinfo{year}{2016}).

\bibitem{Murray2019}
\bibinfo{author}{Murray, C.} \emph{et~al.}
\newblock \bibinfo{title}{{Comprehensive tunneling spectroscopy of
  quasifreestanding MoS$_2$ on graphene on Ir(111)}}.
\newblock \emph{\bibinfo{journal}{Phys. Rev. B}} \textbf{\bibinfo{volume}{99}},
  \bibinfo{pages}{115434} (\bibinfo{year}{2019}).
\newblock \urlprefix\url{https://link.aps.org/doi/10.1103/PhysRevB.99.115434}.

\bibitem{Schuler2019b}
\bibinfo{author}{Schuler, B.} \emph{et~al.}
\newblock \bibinfo{title}{Large spin-orbit splitting of deep in-gap defect
  states of engineered sulfur vacancies in monolayer ${\mathrm{ws}}_{2}$}.
\newblock \emph{\bibinfo{journal}{Phys. Rev. Lett.}}
  \textbf{\bibinfo{volume}{123}}, \bibinfo{pages}{076801}
  (\bibinfo{year}{2019}).
\newblock
  \urlprefix\url{https://link.aps.org/doi/10.1103/PhysRevLett.123.076801}.

\bibitem{Nguyen2019}
\bibinfo{author}{Nguyen, P.~V.} \emph{et~al.}
\newblock \bibinfo{title}{{Visualizing electrostatic gating effects in
  two-dimensional heterostructures}}.
\newblock \emph{\bibinfo{journal}{Nature}} \textbf{\bibinfo{volume}{572}},
  \bibinfo{pages}{220--223} (\bibinfo{year}{2019}).
\newblock \urlprefix\url{http://dx.doi.org/10.1038/s41586-019-1402-1
  http://www.nature.com/articles/s41586-019-1402-1}.
\newblock \eprint{1904.07301}.

\bibitem{Qiu2019}
\bibinfo{author}{Qiu, Z.} \emph{et~al.}
\newblock \bibinfo{title}{{Giant gate-tunable bandgap renormalization and
  excitonic effects in a 2D semiconductor}}.
\newblock \emph{\bibinfo{journal}{Sci. Adv.}} \textbf{\bibinfo{volume}{5}},
  \bibinfo{pages}{2347} (\bibinfo{year}{2019}).
\newblock
  \urlprefix\url{https://advances.sciencemag.org/lookup/doi/10.1126/sciadv.aaw2347}.

\bibitem{Ye2012}
\bibinfo{author}{Ye, J.} \emph{et~al.}
\newblock \bibinfo{title}{Superconducting dome in a gate-tuned band insulator}.
\newblock \emph{\bibinfo{journal}{Science}} \textbf{\bibinfo{volume}{338}},
  \bibinfo{pages}{1193--1196} (\bibinfo{year}{2012}).

\bibitem{Lu2015a}
\bibinfo{author}{Lu, J.~M.} \emph{et~al.}
\newblock \bibinfo{title}{{Evidence for two-dimensional Ising superconductivity
  in gated MoS$_2$}}.
\newblock \emph{\bibinfo{journal}{Science}} \textbf{\bibinfo{volume}{350}},
  \bibinfo{pages}{1353--1357} (\bibinfo{year}{2015}).

\bibitem{Saito2016a}
\bibinfo{author}{Saito, Y.} \emph{et~al.}
\newblock \bibinfo{title}{{Superconductivity protected by spin–valley locking
  in ion-gated MoS$_2$}}.
\newblock \emph{\bibinfo{journal}{Nat. Phys.}} \textbf{\bibinfo{volume}{12}},
  \bibinfo{pages}{144--149} (\bibinfo{year}{2016}).
\newblock \urlprefix\url{http://www.nature.com/articles/nphys3580}.
\newblock \eprint{1506.04146}.

\bibitem{Saito2016}
\bibinfo{author}{Saito, Y.}, \bibinfo{author}{Nojima, T.} \&
  \bibinfo{author}{Iwasa, Y.}
\newblock \bibinfo{title}{{Highly crystalline 2D superconductors}}.
\newblock \emph{\bibinfo{journal}{Nat. Rev. Mater.}}
  \textbf{\bibinfo{volume}{2}}, \bibinfo{pages}{16094} (\bibinfo{year}{2017}).
\newblock \urlprefix\url{http://dx.doi.org/10.1038/natrevmats.2016.94
  http://www.nature.com/articles/natrevmats201694}.

\bibitem{Lu2018}
\bibinfo{author}{Lu, J.} \emph{et~al.}
\newblock \bibinfo{title}{{Full superconducting dome of strong Ising protection
  in gated monolayer WS$_2$}}.
\newblock \emph{\bibinfo{journal}{Proc. Natl. Acad. Sci.}}
  \textbf{\bibinfo{volume}{115}}, \bibinfo{pages}{3551--3556}
  (\bibinfo{year}{2018}).
\newblock
  \urlprefix\url{http://www.pnas.org/lookup/doi/10.1073/pnas.1716781115}.
\newblock \eprint{1703.06369}.

\bibitem{Wang2020c}
\bibinfo{author}{Wang, L.} \emph{et~al.}
\newblock \bibinfo{title}{{Correlated electronic phases in twisted bilayer
  transition metal dichalcogenides}}.
\newblock \emph{\bibinfo{journal}{Nat. Mater.}} \textbf{\bibinfo{volume}{19}},
  \bibinfo{pages}{861--866} (\bibinfo{year}{2020}).
\newblock \urlprefix\url{http://dx.doi.org/10.1038/s41563-020-0708-6
  http://www.nature.com/articles/s41563-020-0708-6}.

\bibitem{Ren2015}
\bibinfo{author}{Ren, Y.} \emph{et~al.}
\newblock \bibinfo{title}{Direct imaging of nanoscale conductance evolution in
  ion-gel-gated oxide transistors}.
\newblock \emph{\bibinfo{journal}{Nano letters}} \textbf{\bibinfo{volume}{15}},
  \bibinfo{pages}{4730--4736} (\bibinfo{year}{2015}).

\bibitem{Jo2015}
\bibinfo{author}{Jo, S.}, \bibinfo{author}{Costanzo, D.},
  \bibinfo{author}{Berger, H.} \& \bibinfo{author}{Morpurgo, A.~F.}
\newblock \bibinfo{title}{{Electrostatically induced superconductivity at the
  surface of WS$_2$}}.
\newblock \emph{\bibinfo{journal}{Nano Lett.}} \textbf{\bibinfo{volume}{15}},
  \bibinfo{pages}{1197--1202} (\bibinfo{year}{2015}).
\newblock \eprint{1501.06531}.

\bibitem{Costanzo2016}
\bibinfo{author}{Costanzo, D.}, \bibinfo{author}{Jo, S.},
  \bibinfo{author}{Berger, H.} \& \bibinfo{author}{Morpurgo, A.~F.}
\newblock \bibinfo{title}{{Gate-induced superconductivity in atomically thin
  MoS$_2$ crystals}}.
\newblock \emph{\bibinfo{journal}{Nat. Nanotechnol.}}
  \textbf{\bibinfo{volume}{11}}, \bibinfo{pages}{339--344}
  (\bibinfo{year}{2016}).
\newblock \urlprefix\url{http://www.nature.com/articles/nnano.2015.314}.
\newblock \eprint{1512.03222}.

\bibitem{Fu2017}
\bibinfo{author}{Fu, Y.} \emph{et~al.}
\newblock \bibinfo{title}{{Gated tuned superconductivity and phonon softening
  in monolayer and bilayer MoS$_2$}}.
\newblock \emph{\bibinfo{journal}{npj Quantum Mater.}}
  \textbf{\bibinfo{volume}{2}}, \bibinfo{pages}{52} (\bibinfo{year}{2017}).
\newblock \urlprefix\url{http://dx.doi.org/10.1038/s41535-017-0056-1
  http://www.nature.com/articles/s41535-017-0056-1}.

\bibitem{Piatti2018}
\bibinfo{author}{Piatti, E.} \emph{et~al.}
\newblock \bibinfo{title}{{Multi-Valley Superconductivity in Ion-Gated MoS$_2$
  Layers}}.
\newblock \emph{\bibinfo{journal}{Nano Lett.}} \textbf{\bibinfo{volume}{18}},
  \bibinfo{pages}{4821--4830} (\bibinfo{year}{2018}).
\newblock
  \urlprefix\url{https://pubs.acs.org/doi/10.1021/acs.nanolett.8b01390}.
\newblock \eprint{1802.06675}.

\bibitem{Costanzo2018}
\bibinfo{author}{Costanzo, D.}, \bibinfo{author}{Zhang, H.},
  \bibinfo{author}{Reddy, B.~A.}, \bibinfo{author}{Berger, H.} \&
  \bibinfo{author}{Morpurgo, A.~F.}
\newblock \bibinfo{title}{{Tunnelling spectroscopy of gate-induced
  superconductivity in MoS$_2$}}.
\newblock \emph{\bibinfo{journal}{Nat. Nanotechnol.}}
  \textbf{\bibinfo{volume}{13}}, \bibinfo{pages}{483--488}
  (\bibinfo{year}{2018}).
\newblock \urlprefix\url{http://dx.doi.org/10.1038/s41565-018-0122-2}.

\bibitem{Helveg2000a}
\bibinfo{author}{Helveg, S.} \emph{et~al.}
\newblock \bibinfo{title}{{Atomic-scale structure of single-layer MoS$_2$
  nanoclusters}}.
\newblock \emph{\bibinfo{journal}{Phys. Rev. Lett.}}
  \textbf{\bibinfo{volume}{84}}, \bibinfo{pages}{951--954}
  (\bibinfo{year}{2000}).
\newblock \urlprefix\url{http://www.ncbi.nlm.nih.gov/pubmed/23959329}.

\bibitem{Alidoust2014}
\bibinfo{author}{Alidoust, N.} \emph{et~al.}
\newblock \bibinfo{title}{{Observation of monolayer valence band spin-orbit
  effect and induced quantum well states in MoX$_2$}}.
\newblock \emph{\bibinfo{journal}{Nat. Commun.}} \textbf{\bibinfo{volume}{5}},
  \bibinfo{pages}{4673} (\bibinfo{year}{2014}).
\newblock \urlprefix\url{http://arxiv.org/abs/1312.7631
  http://www.nature.com/doifinder/10.1038/ncomms5673}.
\newblock \eprint{1312.7631}.

\bibitem{Zhang2014d}
\bibinfo{author}{Zhang, C.}, \bibinfo{author}{Johnson, A.},
  \bibinfo{author}{Hsu, C.-L.}, \bibinfo{author}{Li, L.-J.} \&
  \bibinfo{author}{Shih, C.-K.}
\newblock \bibinfo{title}{{Direct Imaging of Band Profile in Single Layer
  MoS$_2$ on Graphite: Quasiparticle Energy Gap, Metallic Edge States, and Edge
  Band Bending}}.
\newblock \emph{\bibinfo{journal}{Nano Lett.}} \textbf{\bibinfo{volume}{14}},
  \bibinfo{pages}{2443--2447} (\bibinfo{year}{2014}).
\newblock
  \urlprefix\url{http://arxiv.org/abs/1401.5100{\%}5Cnhttp://www.arxiv.org/pdf/1401.5100.pdf
  http://pubs.acs.org/doi/abs/10.1021/nl501133c}.

\bibitem{Kang2017}
\bibinfo{author}{Kang, M.} \emph{et~al.}
\newblock \bibinfo{title}{{Universal Mechanism of Band-Gap Engineering in
  Transition-Metal Dichalcogenides}}.
\newblock \emph{\bibinfo{journal}{Nano Lett.}} \textbf{\bibinfo{volume}{17}},
  \bibinfo{pages}{1610--1615} (\bibinfo{year}{2017}).
\newblock
  \urlprefix\url{https://pubs.acs.org/doi/10.1021/acs.nanolett.6b04775}.

\bibitem{Katoch2018}
\bibinfo{author}{Katoch, J.} \emph{et~al.}
\newblock \bibinfo{title}{{Giant spin-splitting and gap renormalization driven
  by trions in single-layer WS$_2$/h-BN heterostructures}}.
\newblock \emph{\bibinfo{journal}{Nat. Phys.}} \textbf{\bibinfo{volume}{14}},
  \bibinfo{pages}{355--359} (\bibinfo{year}{2018}).
\newblock \urlprefix\url{http://dx.doi.org/10.1038/s41567-017-0033-4
  http://www.nature.com/articles/s41567-017-0033-4}.
\newblock \eprint{1705.04866}.

\bibitem{Liu2020c}
\bibinfo{author}{Liu, H.} \emph{et~al.}
\newblock \bibinfo{title}{{Spontaneous chemical functionalization via
  coordination of Au single atoms on monolayer MoS$_2$}}.
\newblock \emph{\bibinfo{journal}{Sci. Adv.}} \textbf{\bibinfo{volume}{6}},
  \bibinfo{pages}{9308} (\bibinfo{year}{2020}).
\newblock \urlprefix\url{https://advances.sciencemag.org/content/6/49/eabc9308
  https://advances.sciencemag.org/lookup/doi/10.1126/sciadv.abc9308}.

\bibitem{Shao2019}
\bibinfo{author}{Shao, B.} \emph{et~al.}
\newblock \bibinfo{title}{{Pseudodoping of a metallic two-dimensional material
  by the supporting substrate}}.
\newblock \emph{\bibinfo{journal}{Nat. Commun.}} \textbf{\bibinfo{volume}{10}},
  \bibinfo{pages}{180} (\bibinfo{year}{2019}).
\newblock \urlprefix\url{http://dx.doi.org/10.1038/s41467-018-08088-8
  http://www.nature.com/articles/s41467-018-08088-8}.
\newblock \eprint{1807.00756}.

\bibitem{Hsu2017}
\bibinfo{author}{Hsu, Y.-T.}, \bibinfo{author}{Vaezi, A.},
  \bibinfo{author}{Fischer, M.~H.} \& \bibinfo{author}{Kim, E.-A.}
\newblock \bibinfo{title}{{Topological superconductivity in monolayer
  transition metal dichalcogenides}}.
\newblock \emph{\bibinfo{journal}{Nat. Commun.}} \textbf{\bibinfo{volume}{8}},
  \bibinfo{pages}{14985} (\bibinfo{year}{2017}).
\newblock \urlprefix\url{http://www.nature.com/articles/ncomms14985}.
\newblock \eprint{1606.00857}.

\bibitem{Schumacher2013a}
\bibinfo{author}{Schumacher, S.} \emph{et~al.}
\newblock \bibinfo{title}{{The Backside of Graphene: Manipulating Adsorption by
  Intercalation}}.
\newblock \emph{\bibinfo{journal}{Nano Lett.}} \textbf{\bibinfo{volume}{13}},
  \bibinfo{pages}{5013--5019} (\bibinfo{year}{2013}).
\newblock \urlprefix\url{http://www.ncbi.nlm.nih.gov/pubmed/24131290
  http://pubs.acs.org/doi/abs/10.1021/nl402797j
  https://pubs.acs.org/doi/10.1021/nl402797j}.

\bibitem{Larciprete2012}
\bibinfo{author}{Larciprete, R.} \emph{et~al.}
\newblock \bibinfo{title}{{Oxygen Switching of the Epitaxial Graphene–Metal
  Interaction}}.
\newblock \emph{\bibinfo{journal}{ACS Nano}} \textbf{\bibinfo{volume}{6}},
  \bibinfo{pages}{9551--9558} (\bibinfo{year}{2012}).
\newblock \urlprefix\url{http://www.ncbi.nlm.nih.gov/pubmed/23051045
  https://pubs.acs.org/doi/10.1021/nn302729j}.

\bibitem{Granas2012}
\bibinfo{author}{Gr{\aa}n{\"{a}}s, E.} \emph{et~al.}
\newblock \bibinfo{title}{{Oxygen Intercalation under Graphene on Ir(111):
  Energetics, Kinetics, and the Role of Graphene Edges}}.
\newblock \emph{\bibinfo{journal}{ACS Nano}} \textbf{\bibinfo{volume}{6}},
  \bibinfo{pages}{9951--9963} (\bibinfo{year}{2012}).
\newblock \urlprefix\url{http://www.ncbi.nlm.nih.gov/pubmed/23039853
  https://pubs.acs.org/doi/10.1021/nn303548z}.

\bibitem{Jolie2014}
\bibinfo{author}{Jolie, W.} \emph{et~al.}
\newblock \bibinfo{title}{{Confinement of Dirac electrons in graphene quantum
  dots}}.
\newblock \emph{\bibinfo{journal}{Phys. Rev. B}} \textbf{\bibinfo{volume}{89}},
  \bibinfo{pages}{155435} (\bibinfo{year}{2014}).
\newblock \urlprefix\url{http://link.aps.org/doi/10.1103/PhysRevB.89.155435}.

\bibitem{Martinez-Galera2016}
\bibinfo{author}{Mart{\'{i}}nez-Galera, A.~J.} \emph{et~al.}
\newblock \bibinfo{title}{{Oxygen orders differently under graphene: new
  superstructures on Ir(111)}}.
\newblock \emph{\bibinfo{journal}{Nanoscale}} \textbf{\bibinfo{volume}{8}},
  \bibinfo{pages}{1932--1943} (\bibinfo{year}{2016}).
\newblock \urlprefix\url{http://xlink.rsc.org/?DOI=C5NR04976H}.

\bibitem{Schumacher2013}
\bibinfo{author}{Schumacher, S.}, \bibinfo{author}{F{\"{o}}rster, D.},
  \bibinfo{author}{R{\"{o}}sner, M.}, \bibinfo{author}{Wehling, T.} \&
  \bibinfo{author}{Michely, T.}
\newblock \bibinfo{title}{{Strain in Epitaxial Graphene Visualized by
  Intercalation}}.
\newblock \emph{\bibinfo{journal}{Phys. Rev. Lett.}}
  \textbf{\bibinfo{volume}{110}}, \bibinfo{pages}{86111}
  (\bibinfo{year}{2013}).
\newblock
  \urlprefix\url{http://link.aps.org/doi/10.1103/PhysRevLett.110.086111}.

\bibitem{Huttmann2015}
\bibinfo{author}{Huttmann, F.} \emph{et~al.}
\newblock \bibinfo{title}{{Tuning the van der Waals Interaction of Graphene
  with Molecules via Doping}}.
\newblock \emph{\bibinfo{journal}{Phys. Rev. Lett.}}
  \textbf{\bibinfo{volume}{115}}, \bibinfo{pages}{236101}
  (\bibinfo{year}{2015}).
\newblock
  \urlprefix\url{https://link.aps.org/doi/10.1103/PhysRevLett.115.236101}.
\newblock \eprint{1509.02020}.

\bibitem{Zhang2015}
\bibinfo{author}{Zhang, X.} \emph{et~al.}
\newblock \bibinfo{title}{{Growth graphene on silver–copper nanoparticles by
  chemical vapor deposition for high-performance surface-enhanced Raman
  scattering}}.
\newblock \emph{\bibinfo{journal}{Appl. Surf. Sci.}}
  \textbf{\bibinfo{volume}{353}}, \bibinfo{pages}{63--70}
  (\bibinfo{year}{2015}).
\newblock
  \urlprefix\url{http://linkinghub.elsevier.com/retrieve/pii/S0169433215014282}.

\bibitem{Ehlen2019}
\bibinfo{author}{Ehlen, N.} \emph{et~al.}
\newblock \bibinfo{title}{{Narrow photoluminescence peak of epitaxial MoS$_2$
  on graphene/Ir(111)}}.
\newblock \emph{\bibinfo{journal}{2D Mater.}} \textbf{\bibinfo{volume}{6}},
  \bibinfo{pages}{011006} (\bibinfo{year}{2019}).
\newblock \urlprefix\url{http://arxiv.org/abs/1809.01886}.
\newblock \eprint{1809.01886}.

\bibitem{Ehlen2017}
\bibinfo{author}{{N. Ehlen}}.
\newblock \bibinfo{title}{{Private Communication}}.

\bibitem{Komsa2013_PRB}
\bibinfo{author}{Komsa, H.-P.} \& \bibinfo{author}{Krasheninnikov, A.~V.}
\newblock \bibinfo{title}{Electronic structures and optical properties of
  realistic transition metal dichalcogenide heterostructures from first
  principles}.
\newblock \emph{\bibinfo{journal}{Phys. Rev. B}} \textbf{\bibinfo{volume}{88}},
  \bibinfo{pages}{085318} (\bibinfo{year}{2013}).
\newblock \urlprefix\url{http://link.aps.org/doi/10.1103/PhysRevB.88.085318}.

\bibitem{Li1985}
\bibinfo{author}{Li, W.-K.} \& \bibinfo{author}{Blinder, S.~M.}
\newblock \bibinfo{title}{Solution of the schr{\"o}dinger equation for a
  particle in an equilateral triangle}.
\newblock \emph{\bibinfo{journal}{J. Math. Phys.}}
  \textbf{\bibinfo{volume}{26}}, \bibinfo{pages}{2784--2786}
  (\bibinfo{year}{1985}).

\bibitem{Cheiwchanchamnangij2012}
\bibinfo{author}{Cheiwchanchamnangij, T.} \& \bibinfo{author}{Lambrecht, W.
  R.~L.}
\newblock \bibinfo{title}{{Quasiparticle band structure calculation of
  monolayer, bilayer, and bulk MoS$_2$}}.
\newblock \emph{\bibinfo{journal}{Phys. Rev. B}} \textbf{\bibinfo{volume}{85}},
  \bibinfo{pages}{205302} (\bibinfo{year}{2012}).

\bibitem{Qiu2013a}
\bibinfo{author}{Qiu, D.~Y.}, \bibinfo{author}{da~Jornada, F.~H.} \&
  \bibinfo{author}{Louie, S.~G.}
\newblock \bibinfo{title}{{Optical Spectrum of MoS$_2$: Many-Body Effects and
  Diversity of Exciton States}}.
\newblock \emph{\bibinfo{journal}{Phys. Rev. Lett.}}
  \textbf{\bibinfo{volume}{111}}, \bibinfo{pages}{216805}
  (\bibinfo{year}{2013}).
\newblock
  \urlprefix\url{https://link.aps.org/doi/10.1103/PhysRevLett.111.216805}.
\newblock \eprint{arXiv:1311.0963v1}.

\bibitem{Komsa2015}
\bibinfo{author}{Komsa, H.-P.} \& \bibinfo{author}{Krasheninnikov, A.~V.}
\newblock \bibinfo{title}{Native defects in bulk and monolayer
  ${\mathrm{mos}}_{2}$ from first principles}.
\newblock \emph{\bibinfo{journal}{Phys. Rev. B}} \textbf{\bibinfo{volume}{91}},
  \bibinfo{pages}{125304} (\bibinfo{year}{2015}).
\newblock \urlprefix\url{https://link.aps.org/doi/10.1103/PhysRevB.91.125304}.

\bibitem{Michaelson1977}
\bibinfo{author}{Michaelson, H.~B.}
\newblock \bibinfo{title}{The work function of the elements and its
  periodicity}.
\newblock \emph{\bibinfo{journal}{J. Appl. Phys.}}
  \textbf{\bibinfo{volume}{48}}, \bibinfo{pages}{4729--4733}
  (\bibinfo{year}{1977}).

\bibitem{Yan2012}
\bibinfo{author}{Yan, H.}, \bibinfo{author}{Sun, Y.}, \bibinfo{author}{He, L.},
  \bibinfo{author}{Nie, J.-C.} \& \bibinfo{author}{Chan, M. H.~W.}
\newblock \bibinfo{title}{{Observation of Landau-level-like quantization at 77
  K along a strained-induced graphene ridge}}.
\newblock \emph{\bibinfo{journal}{Phys. Rev. B}} \textbf{\bibinfo{volume}{85}},
  \bibinfo{pages}{035422} (\bibinfo{year}{2012}).
\newblock \urlprefix\url{http://link.aps.org/doi/10.1103/PhysRevB.85.035422}.

\bibitem{Pletikosic2009}
\bibinfo{author}{Pletikosi{\'{c}}, I.} \emph{et~al.}
\newblock \bibinfo{title}{{Dirac Cones and Minigaps for Graphene on Ir(111)}}.
\newblock \emph{\bibinfo{journal}{Phys. Rev. Lett.}}
  \textbf{\bibinfo{volume}{102}}, \bibinfo{pages}{056808}
  (\bibinfo{year}{2009}).
\newblock \urlprefix\url{http://link.aps.org/doi/10.1103/PhysRevLett.102.056808
  https://link.aps.org/doi/10.1103/PhysRevLett.102.056808}.

\bibitem{Khomyakov2009}
\bibinfo{author}{Khomyakov, P.~A.} \emph{et~al.}
\newblock \bibinfo{title}{{First-principles study of the interaction and charge
  transfer between graphene and metals}}.
\newblock \emph{\bibinfo{journal}{Phys. Rev. B}} \textbf{\bibinfo{volume}{79}},
  \bibinfo{pages}{195425} (\bibinfo{year}{2009}).
\newblock \urlprefix\url{http://link.aps.org/doi/10.1103/PhysRevB.79.195425}.

\bibitem{Das2013}
\bibinfo{author}{Das, S.}, \bibinfo{author}{Chen, H.-Y.},
  \bibinfo{author}{Penumatcha, A.~V.} \& \bibinfo{author}{Appenzeller, J.}
\newblock \bibinfo{title}{{High performance multilayer MoS$_2$ transistors with
  scandium contacts}}.
\newblock \emph{\bibinfo{journal}{Nano Lett.}} \textbf{\bibinfo{volume}{13}},
  \bibinfo{pages}{100--105} (\bibinfo{year}{2013}).

\bibitem{Lee2016d}
\bibinfo{author}{Lee, S.~Y.} \emph{et~al.}
\newblock \bibinfo{title}{{Large work function modulation of monolayer MoS$_2$
  by ambient gases}}.
\newblock \emph{\bibinfo{journal}{Acs Nano}} \textbf{\bibinfo{volume}{10}},
  \bibinfo{pages}{6100--6107} (\bibinfo{year}{2016}).

\bibitem{Hu2018}
\bibinfo{author}{Hu, C.} \emph{et~al.}
\newblock \bibinfo{title}{{Work function variation of monolayer MoS$_2$ by
  nitrogen-doping}}.
\newblock \emph{\bibinfo{journal}{Appl. Phys. Lett.}}
  \textbf{\bibinfo{volume}{113}}, \bibinfo{pages}{041602}
  (\bibinfo{year}{2018}).

\bibitem{Zhang2016a}
\bibinfo{author}{Zhang, C.} \emph{et~al.}
\newblock \bibinfo{title}{Systematic study of electronic structure and band
  alignment of monolayer transition metal dichalcogenides in van der waals
  heterostructures}.
\newblock \emph{\bibinfo{journal}{2D Mater.}} \textbf{\bibinfo{volume}{4}},
  \bibinfo{pages}{015026} (\bibinfo{year}{2016}).

\bibitem{Schlaf1999}
\bibinfo{author}{Schlaf, R.}, \bibinfo{author}{Lang, O.},
  \bibinfo{author}{Pettenkofer, C.} \& \bibinfo{author}{Jaegermann, W.}
\newblock \bibinfo{title}{{Band lineup of layered semiconductor
  heterointerfaces prepared by van der Waals epitaxy: Charge transfer
  correction term for the electron affinity rule}}.
\newblock \emph{\bibinfo{journal}{J. Appl. Phys.}}
  \textbf{\bibinfo{volume}{85}}, \bibinfo{pages}{2732} (\bibinfo{year}{1999}).
\newblock
  \urlprefix\url{http://scitation.aip.org/content/aip/journal/jap/85/5/10.1063/1.369590{\%}5Cnhttp://scitation.aip.org/deliver/fulltext/aip/journal/jap/85/5/1.369590.pdf;jsessionid=14gka73dl0h9a.x-aip-live-06?itemId=/content/aip/journal/jap/85/5/10.1063/1.369590{\&}mimeType=pdf}.

\bibitem{Riis-Jensen2020}
\bibinfo{author}{Riis-Jensen, A.~C.}, \bibinfo{author}{Lu, J.} \&
  \bibinfo{author}{Thygesen, K.~S.}
\newblock \bibinfo{title}{Electrically controlled dielectric band gap
  engineering in a two-dimensional semiconductor}.
\newblock \emph{\bibinfo{journal}{Phys. Rev. B}}
  \textbf{\bibinfo{volume}{101}}, \bibinfo{pages}{121110}
  (\bibinfo{year}{2020}).

\bibitem{Ugeda2014}
\bibinfo{author}{Ugeda, M.~M.} \emph{et~al.}
\newblock \bibinfo{title}{{Giant bandgap renormalization and excitonic effects
  in a monolayer transition metal dichalcogenide semiconductor}}.
\newblock \emph{\bibinfo{journal}{Nat. Mater.}} \textbf{\bibinfo{volume}{13}},
  \bibinfo{pages}{1091--1095} (\bibinfo{year}{2014}).
\newblock \urlprefix\url{http://www.nature.com/doifinder/10.1038/nmat4061}.

\bibitem{Allain2015}
\bibinfo{author}{Allain, A.}, \bibinfo{author}{Kang, J.},
  \bibinfo{author}{Banerjee, K.} \& \bibinfo{author}{Kis, A.}
\newblock \bibinfo{title}{Electrical contacts to two-dimensional
  semiconductors}.
\newblock \emph{\bibinfo{journal}{Nature materials}}
  \textbf{\bibinfo{volume}{14}}, \bibinfo{pages}{1195--1205}
  (\bibinfo{year}{2015}).

\bibitem{Diaz2016}
\bibinfo{author}{Diaz, H.~C.}, \bibinfo{author}{Ma, Y.},
  \bibinfo{author}{Chaghi, R.} \& \bibinfo{author}{Batzill, M.}
\newblock \bibinfo{title}{{High density of (pseudo) periodic twin-grain
  boundaries in molecular beam epitaxy-grown van der Waals heterostructure:
  MoTe$_2$/MoS$_2$}}.
\newblock \emph{\bibinfo{journal}{Appl. Phys. Lett.}}
  \textbf{\bibinfo{volume}{108}}, \bibinfo{pages}{191606}
  (\bibinfo{year}{2016}).
\newblock \urlprefix\url{http://aip.scitation.org/doi/10.1063/1.4949559}.

\bibitem{Bampoulis2017}
\bibinfo{author}{Bampoulis, P.} \emph{et~al.}
\newblock \bibinfo{title}{{Defect Dominated Charge Transport and Fermi Level
  Pinning in MoS$_2$/Metal Contacts}}.
\newblock \emph{\bibinfo{journal}{ACS Applied Materials \& Interfaces}}
  \textbf{\bibinfo{volume}{9}}, \bibinfo{pages}{19278--19286}
  (\bibinfo{year}{2017}).
\newblock \urlprefix\url{https://doi.org/10.1021/acsami.7b02739}.
\newblock \bibinfo{note}{PMID: 28508628},
  \eprint{https://doi.org/10.1021/acsami.7b02739}.

\bibitem{LeQuang2017}
\bibinfo{author}{{Le Quang}, T.} \emph{et~al.}
\newblock \bibinfo{title}{{Scanning tunneling spectroscopy of van der Waals
  graphene/semiconductor interfaces: absence of Fermi level pinning}}.
\newblock \emph{\bibinfo{journal}{2D Materials}} \textbf{\bibinfo{volume}{4}},
  \bibinfo{pages}{035019} (\bibinfo{year}{2017}).
\newblock \urlprefix\url{https://doi.org/10.1088/2053-1583/aa7b03}.

\bibitem{Murray2020}
\bibinfo{author}{Murray, C.} \emph{et~al.}
\newblock \bibinfo{title}{{Band Bending and Valence Band Quantization at Line
  Defects in MoS$_2$}}.
\newblock \emph{\bibinfo{journal}{ACS Nano}} \textbf{\bibinfo{volume}{14}},
  \bibinfo{pages}{9176--9187} (\bibinfo{year}{2020}).
\newblock \urlprefix\url{https://pubs.acs.org/doi/10.1021/acsnano.0c04945}.

\bibitem{Zou2013}
\bibinfo{author}{Zou, X.}, \bibinfo{author}{Liu, Y.} \&
  \bibinfo{author}{Yakobson, B.~I.}
\newblock \bibinfo{title}{{Predicting Dislocations and Grain Boundaries in
  Two-Dimensional Metal-Disulfides from the First Principles}}.
\newblock \emph{\bibinfo{journal}{Nano Lett.}} \textbf{\bibinfo{volume}{13}},
  \bibinfo{pages}{253--258} (\bibinfo{year}{2013}).
\newblock \urlprefix\url{http://pubs.acs.org/doi/abs/10.1021/nl3040042}.

\bibitem{Komsa2017}
\bibinfo{author}{Komsa, H.-P.} \& \bibinfo{author}{Krasheninnikov, A.~V.}
\newblock \bibinfo{title}{{Engineering the Electronic Properties of
  Two-Dimensional Transition Metal Dichalcogenides by Introducing Mirror Twin
  Boundaries}}.
\newblock \emph{\bibinfo{journal}{Adv. Electron. Mater.}}
  \textbf{\bibinfo{volume}{3}}, \bibinfo{pages}{1600468}
  (\bibinfo{year}{2017}).
\newblock \urlprefix\url{http://doi.wiley.com/10.1002/aelm.201600468}.

\bibitem{Jolie2019a}
\bibinfo{author}{Jolie, W.} \emph{et~al.}
\newblock \bibinfo{title}{{Tomonaga-Luttinger Liquid in a Box: Electrons
  Confined within MoS$_2$ Mirror-Twin Boundaries}}.
\newblock \emph{\bibinfo{journal}{Phys. Rev. X}} \textbf{\bibinfo{volume}{9}},
  \bibinfo{pages}{11055} (\bibinfo{year}{2019}).
\newblock \urlprefix\url{https://doi.org/10.1103/PhysRevX.9.011055}.

\bibitem{Kormanyos2015}
\bibinfo{author}{Korm{\'{a}}nyos, A.} \emph{et~al.}
\newblock \bibinfo{title}{k $\cdotp$ p theory for two-dimensional transition
  metal dichalcogenide semiconductors}.
\newblock \emph{\bibinfo{journal}{2D Mater.}} \textbf{\bibinfo{volume}{2}},
  \bibinfo{pages}{022001} (\bibinfo{year}{2015}).
\newblock \urlprefix\url{https://doi.org/10.1088/2053-1583/2/2/022001}.

\bibitem{Yang2021a}
\bibinfo{author}{Yang, X.} \emph{et~al.}
\newblock \bibinfo{title}{Manipulating hubbard-type coulomb blockade effect of
  metallic wires embedded in an insulator}.
\newblock \emph{\bibinfo{journal}{arXiv preprint arXiv:2104.08577}}
  (\bibinfo{year}{2021}).

\bibitem{Petrovic2013}
\bibinfo{author}{Petrovi{\'{c}}, M.} \emph{et~al.}
\newblock \bibinfo{title}{{The mechanism of caesium intercalation of
  graphene}}.
\newblock \emph{\bibinfo{journal}{Nat. Commun.}} \textbf{\bibinfo{volume}{4}},
  \bibinfo{pages}{2772} (\bibinfo{year}{2013}).
\newblock \urlprefix\url{http://www.ncbi.nlm.nih.gov/pubmed/24212475
  http://www.nature.com/doifinder/10.1038/ncomms3772
  http://www.nature.com/articles/ncomms3772}.

\bibitem{Klimovskikh2019}
\bibinfo{author}{Klimovskikh, I.}, \bibinfo{author}{Krivenkov, M.},
  \bibinfo{author}{Varykhalov, A.}, \bibinfo{author}{Estyunin, D.} \&
  \bibinfo{author}{Shikin, A.}
\newblock \bibinfo{title}{Reconstructed fermi surface in graphene on ir(111) by
  gd-ir surface alloying}.
\newblock \emph{\bibinfo{journal}{Carbon}} \textbf{\bibinfo{volume}{147}},
  \bibinfo{pages}{182--186} (\bibinfo{year}{2019}).

\bibitem{Vinogradov2012}
\bibinfo{author}{Vinogradov, N.~A.} \emph{et~al.}
\newblock \bibinfo{title}{Controllable p-doping of graphene on ir (111) by
  chlorination with fecl$_3$}.
\newblock \emph{\bibinfo{journal}{J. Phys. Condens. Matter}}
  \textbf{\bibinfo{volume}{24}}, \bibinfo{pages}{314202}
  (\bibinfo{year}{2012}).

\bibitem{Schroeder2016}
\bibinfo{author}{Schr{\"{o}}der, U.~A.} \emph{et~al.}
\newblock \bibinfo{title}{{Core level shifts of intercalated graphene}}.
\newblock \emph{\bibinfo{journal}{2D Mater.}} \textbf{\bibinfo{volume}{4}},
  \bibinfo{pages}{015013} (\bibinfo{year}{2016}).
\newblock
  \urlprefix\url{http://stacks.iop.org/2053-1583/4/i=1/a=015013?key=crossref.d48154c3b537a0e53e906f8854852f6d
  https://iopscience.iop.org/article/10.1088/2053-1583/4/1/015013}.

\bibitem{Halle2016}
\bibinfo{author}{Halle, J.}, \bibinfo{author}{N{\'{e}}el, N.} \&
  \bibinfo{author}{Kr{\"{o}}ger, J.}
\newblock \bibinfo{title}{{Filling the Gap: Li-Intercalated Graphene on
  Ir(111)}}.
\newblock \emph{\bibinfo{journal}{J. Phys. Chem. C}}
  \textbf{\bibinfo{volume}{120}}, \bibinfo{pages}{5067--5073}
  (\bibinfo{year}{2016}).
\newblock \urlprefix\url{http://pubs.acs.org/doi/abs/10.1021/acs.jpcc.6b00729}.

\bibitem{Coraux2009}
\bibinfo{author}{Coraux, J.} \emph{et~al.}
\newblock \bibinfo{title}{{Growth of graphene on Ir(111)}}.
\newblock \emph{\bibinfo{journal}{New J. Phys.}} \textbf{\bibinfo{volume}{11}},
  \bibinfo{pages}{023006} (\bibinfo{year}{2009}).
\newblock
  \urlprefix\url{http://stacks.iop.org/1367-2630/11/i=2/a=023006?key=crossref.41fefc8129b7d1091023ff6f80606e2b}.

\bibitem{Hall2017}
\bibinfo{author}{Hall, J.} \emph{et~al.}
\newblock \bibinfo{title}{{Molecular beam epitaxy of quasi-freestanding
  transition metal disulphide monolayers on van der Waals substrates: a growth
  study}}.
\newblock \emph{\bibinfo{journal}{2D Mater.}} \textbf{\bibinfo{volume}{5}},
  \bibinfo{pages}{025005} (\bibinfo{year}{2018}).
\newblock
  \urlprefix\url{http://stacks.iop.org/2053-1583/5/i=2/a=025005?key=crossref.ccfb46f08c689ba275c96d24f12bc040}.

\bibitem{Schumacher2014}
\bibinfo{author}{Schumacher, S.} \emph{et~al.}
\newblock \bibinfo{title}{{Europium underneath graphene on Ir(111):
  Intercalation mechanism, magnetism, and band structure}}.
\newblock \emph{\bibinfo{journal}{Phys. Rev. B}} \textbf{\bibinfo{volume}{90}},
  \bibinfo{pages}{235437} (\bibinfo{year}{2014}).
\newblock \urlprefix\url{https://link.aps.org/doi/10.1103/PhysRevB.90.235437}.
\newblock \eprint{1409.4272}.

\bibitem{Morgenstern2003}
\bibinfo{author}{Morgenstern, M.}
\newblock \bibinfo{title}{{Probing the local density of states of dilute
  electron systems in different dimensions}}.
\newblock \emph{\bibinfo{journal}{Surf. Rev. Lett.}}
  \textbf{\bibinfo{volume}{10}}, \bibinfo{pages}{933--962}
  (\bibinfo{year}{2003}).
\newblock
  \urlprefix\url{http://www.worldscientific.com/doi/abs/10.1142/S0218625X0300575X}.

\bibitem{vasp1}
\bibinfo{author}{Kresse, G.} \& \bibinfo{author}{Furthm\"uller, J.}
\newblock \bibinfo{title}{Efficient iterative schemes for ab initio
  total-energy calculations using a plane-wave basis set}.
\newblock \emph{\bibinfo{journal}{Phys. Rev. B}} \textbf{\bibinfo{volume}{54}},
  \bibinfo{pages}{11169--11186} (\bibinfo{year}{1996}).

\bibitem{vasp2}
\bibinfo{author}{Kresse, G.} \& \bibinfo{author}{Furthm\"uller, J.}
\newblock \bibinfo{title}{Efficiency of ab-initio total energy calculations for
  metals and semiconductors using a plane-wave basis set}.
\newblock \emph{\bibinfo{journal}{Comp. Mater. Sci.}}
  \textbf{\bibinfo{volume}{6}}, \bibinfo{pages}{15--50} (\bibinfo{year}{1996}).

\bibitem{PBE}
\bibinfo{author}{Perdew, J.~P.}, \bibinfo{author}{Burke, K.} \&
  \bibinfo{author}{Ernzerhof, M.}
\newblock \bibinfo{title}{Generalized gradient approximation made simple}.
\newblock \emph{\bibinfo{journal}{Phys. Rev. Lett.}}
  \textbf{\bibinfo{volume}{77}}, \bibinfo{pages}{3865} (\bibinfo{year}{1996}).

\bibitem{GrimmeD2}
\bibinfo{author}{Grimme, S.}
\newblock \bibinfo{title}{Semiempirical gga-type density functional constructed
  with a long-range dispersion correction}.
\newblock \emph{\bibinfo{journal}{J. Comput. Chem.}}
  \textbf{\bibinfo{volume}{27}}, \bibinfo{pages}{1787} (\bibinfo{year}{2006}).

\end{thebibliography}


\begin{thebibliography}{10}
\expandafter\ifx\csname url\endcsname\relax
  \def\url#1{\texttt{#1}}\fi
\expandafter\ifx\csname urlprefix\endcsname\relax\def\urlprefix{URL }\fi
\providecommand{\bibinfo}[2]{#2}
\providecommand{\eprint}[2][]{\url{#2}}

\bibitem{Foerster2012}
\bibinfo{author}{F{\"o}rster, D.~F.}, \bibinfo{author}{Wehling, T.~O.},
  \bibinfo{author}{Schumacher, S.}, \bibinfo{author}{Rosch, A.} \&
  \bibinfo{author}{Michely, T.}
\newblock \bibinfo{title}{Phase coexistence of clusters and islands: Europium
  on graphene}.
\newblock \emph{\bibinfo{journal}{New Journal of Physics}}
  \textbf{\bibinfo{volume}{14}}, \bibinfo{pages}{023022}
  (\bibinfo{year}{2012}).

\bibitem{Schumacher2013b}
\bibinfo{author}{Schumacher, S.}, \bibinfo{author}{F{\"{o}}rster, D.~F.},
  \bibinfo{author}{R{\"{o}}sner, M.}, \bibinfo{author}{Wehling, T.~O.} \&
  \bibinfo{author}{Michely, T.}
\newblock \bibinfo{title}{{Strain in epitaxial graphene visualized by
  intercalation}}.
\newblock \emph{\bibinfo{journal}{Phys. Rev. Lett.}}
  \textbf{\bibinfo{volume}{110}}, \bibinfo{pages}{86111}
  (\bibinfo{year}{2013}).

\bibitem{Schumacher2014}
\bibinfo{author}{Schumacher, S.} \emph{et~al.}
\newblock \bibinfo{title}{{Europium underneath graphene on Ir(111):
  Intercalation mechanism, magnetism, and band structure}}.
\newblock \emph{\bibinfo{journal}{Phys. Rev. B}} \textbf{\bibinfo{volume}{90}},
  \bibinfo{pages}{235437} (\bibinfo{year}{2014}).
\newblock \urlprefix\url{https://link.aps.org/doi/10.1103/PhysRevB.90.235437}.
\newblock \eprint{1409.4272}.

\bibitem{Martinez-Galera2016}
\bibinfo{author}{Mart{\'{i}}nez-Galera, A.~J.} \emph{et~al.}
\newblock \bibinfo{title}{{Oxygen orders differently under graphene: new
  superstructures on Ir(111)}}.
\newblock \emph{\bibinfo{journal}{Nanoscale}} \textbf{\bibinfo{volume}{8}},
  \bibinfo{pages}{1932--1943} (\bibinfo{year}{2016}).
\newblock \urlprefix\url{http://xlink.rsc.org/?DOI=C5NR04976H}.

\bibitem{Crommie1993b}
\bibinfo{author}{Crommie, M.~F.}, \bibinfo{author}{Lutz, C.~P.} \&
  \bibinfo{author}{Eigler, D.~M.}
\newblock \bibinfo{title}{{Imaging standing waves in a two-dimensional electron
  gas}}.
\newblock \emph{\bibinfo{journal}{Nature}} \textbf{\bibinfo{volume}{363}},
  \bibinfo{pages}{524--527} (\bibinfo{year}{1993}).
\newblock \urlprefix\url{http://www.nature.com/doifinder/10.1038/363524a0}.

\bibitem{Li1985}
\bibinfo{author}{Li, W.-K.} \& \bibinfo{author}{Blinder, S.~M.}
\newblock \bibinfo{title}{Solution of the schr{\"o}dinger equation for a
  particle in an equilateral triangle}.
\newblock \emph{\bibinfo{journal}{J. Math. Phys.}}
  \textbf{\bibinfo{volume}{26}}, \bibinfo{pages}{2784--2786}
  (\bibinfo{year}{1985}).

\bibitem{Jolie2014}
\bibinfo{author}{Jolie, W.} \emph{et~al.}
\newblock \bibinfo{title}{{Confinement of Dirac electrons in graphene quantum
  dots}}.
\newblock \emph{\bibinfo{journal}{Phys. Rev. B}} \textbf{\bibinfo{volume}{89}},
  \bibinfo{pages}{155435} (\bibinfo{year}{2014}).
\newblock \urlprefix\url{http://link.aps.org/doi/10.1103/PhysRevB.89.155435}.

\bibitem{Jolie2015}
\bibinfo{author}{Jolie, W.}, \bibinfo{author}{Craes, F.} \&
  \bibinfo{author}{Busse, C.}
\newblock \bibinfo{title}{{Graphene on weakly interacting metals: Dirac states
  versus surface states}}.
\newblock \emph{\bibinfo{journal}{Phys. Rev. B}} \textbf{\bibinfo{volume}{91}},
  \bibinfo{pages}{115419} (\bibinfo{year}{2015}).
\newblock \urlprefix\url{https://link.aps.org/doi/10.1103/PhysRevB.91.115419}.

\bibitem{Zhang2008}
\bibinfo{author}{Zhang, Y.} \emph{et~al.}
\newblock \bibinfo{title}{{Giant phonon-induced conductance in scanning
  tunnelling spectroscopy of gate-tunable graphene}}.
\newblock \emph{\bibinfo{journal}{Nat. Phys.}} \textbf{\bibinfo{volume}{4}},
  \bibinfo{pages}{627--630} (\bibinfo{year}{2008}).
\newblock \urlprefix\url{http://www.nature.com/doifinder/10.1038/nphys1022}.

\bibitem{Natterer2015}
\bibinfo{author}{Natterer, F.~D.} \emph{et~al.}
\newblock \bibinfo{title}{{Strong Asymmetric Charge Carrier Dependence in
  Inelastic Electron Tunneling Spectroscopy of Graphene Phonons}}.
\newblock \emph{\bibinfo{journal}{Phys. Rev. Lett.}}
  \textbf{\bibinfo{volume}{114}}, \bibinfo{pages}{245502}
  (\bibinfo{year}{2015}).
\newblock
  \urlprefix\url{http://link.aps.org/doi/10.1103/PhysRevLett.114.245502}.

\bibitem{Lin2007}
\bibinfo{author}{Lin, C.~L.} \emph{et~al.}
\newblock \bibinfo{title}{Manifestation of work function difference in high
  order gundlach oscillation}.
\newblock \emph{\bibinfo{journal}{Phys. Rev. Lett.}}
  \textbf{\bibinfo{volume}{99}}, \bibinfo{pages}{216103}
  (\bibinfo{year}{2007}).
\newblock
  \urlprefix\url{https://link.aps.org/doi/10.1103/PhysRevLett.99.216103}.

\bibitem{Pletikosic2009}
\bibinfo{author}{Pletikosi{\'{c}}, I.} \emph{et~al.}
\newblock \bibinfo{title}{{Dirac Cones and Minigaps for Graphene on Ir(111)}}.
\newblock \emph{\bibinfo{journal}{Phys. Rev. Lett.}}
  \textbf{\bibinfo{volume}{102}}, \bibinfo{pages}{056808}
  (\bibinfo{year}{2009}).
\newblock \urlprefix\url{http://link.aps.org/doi/10.1103/PhysRevLett.102.056808
  https://link.aps.org/doi/10.1103/PhysRevLett.102.056808}.

\bibitem{Murray2020}
\bibinfo{author}{Murray, C.} \emph{et~al.}
\newblock \bibinfo{title}{{Band Bending and Valence Band Quantization at Line
  Defects in MoS$_2$}}.
\newblock \emph{\bibinfo{journal}{ACS Nano}} \textbf{\bibinfo{volume}{14}},
  \bibinfo{pages}{9176--9187} (\bibinfo{year}{2020}).
\newblock \urlprefix\url{https://pubs.acs.org/doi/10.1021/acsnano.0c04945}.

\end{thebibliography}

\end{document}

\title{Supplementary Information of `Metal-insulator transition in monolayer \mo{} via contactless chemical doping'}
\maketitle
\author{Camiel van Efferen$^{1,5}*$,}
\author{Clifford Murray$^{1,5}$,}
\author{Jeison Fischer$^1$,} 
\author{Carsten Busse$^{2,3}$,}
\author{Hannu-Pekka Komsa$^4$,}
\author{Thomas Michely$^1$,}
\author{Wouter Jolie$^{1}$} 
\begin{affiliations}
\item II. Physikalisches Institut, Universit\"{a}t zu K\"{o}ln, Z\"{u}lpicher Stra\ss e 77, 50937 K\"{o}ln, Germany
\item Present address: Department Physik, Universität Siegen, Walter-Flex-Str. 3, 57068 Siegen, Germany
\item Institut f{\"ur} Materialphysik, Westf\"{a}lische Wilhelms-Universit\"{a}t M\"{u}nster, Wilhelm-Klemm-Stra{\ss}e 10, 48149 M\"{u}nster, Germany
\item Faculty of Information Technology and Electrical Engineering, University of Oulu, Pentti Kaiteran katu 1, 90014 Oulu, Finland
\item CvE and CM contributed equally to this work
\end{affiliations}
	
\clearpage

\begin{figure*}[h!]
	\centering
	\includegraphics[width=\columnwidth]{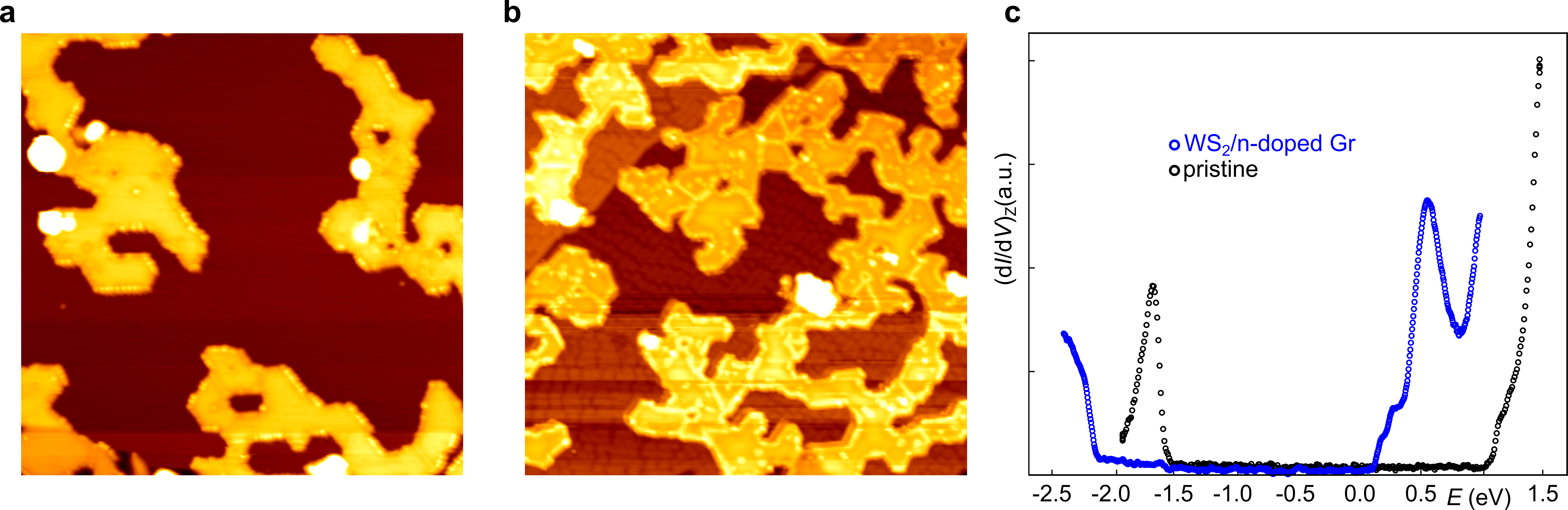}
	\caption{\footnotesize \textbf{a}~Constant current STM overview topograph of WS$_2$/Gr/Ir(111) Sample has 0.4 ML coverage of monolayer WS$_2$ with few bilayers. \textbf{b}~WS$_2$/Gr/Eu/Ir(111). Sample has 0.7 ML coverage of monolayer WS$_2$ with no bilayers present. A single Ir(111) step edge is visible in the upper left. Gr is fully intercalated. \textbf{c}~Constant height STS of ML WS$_2$/Gr/Ir(111) (black) and ML WS$_2$/Gr/Eu/Ir(111) (blue).
		STM/STS parameters:
			\textbf{a}~$V_\text{st}$= \SI{1.75}{\V}, $I_\text{st}$=\SI{0.03}{\nA}, image size \SI{100 \times 100}{\nano\meter\squared};
			\textbf{b}~$V_\text{st}$= \SI{-2.0}{\V}, $I_\text{st}$=\SI{0.1}{\nA}, image size \SI{100 \times 100}{\nano\meter\squared};
			\textbf{c}~$V_\text{st}$= \SI{1.5}{\V}, $I_\text{st}$=\SI{0.4}{\nA} (black); $V_\text{st}$= \SI{-2.5}{\V}, $I_\text{st}$=\SI{0.2}{\nA} (blue).
			\label{supp:figWS2}}
	\end{figure*}

\section*{Fig.~\ref{supp:figWS2}: Contactless chemical doping of WS$_2$}

Contactless chemical gating of WS$_2$ is demonstrated in Fig.~\ref{supp:figWS2}. The scanning tunneling microscopy (STM) topographs, displayed in Fig.~\ref{supp:figWS2}a, b, present ML WS$_2$ islands on graphene (Gr) and on n-gated Gr, respectively. In the latter, Gr is fully intercalated by Eu. The Eu superstructure is identical to the one found after \mo{} growth. It is  discussed in Fig.\,\ref{supp:STMtopo}. 

Fig.~\ref{supp:figWS2}b compares constant height spectra measured on WS$_2$/Gr/Ir(111) (referred to as pristine) and WS$_2$/Gr/Eu/Ir(111) (referred to as WS$_2$~on n-doped graphene). We see that the n-doping of Gr causes a considerable shift in the WS$_2$ band structure towards lower energies. The conduction band (CB) is shifted by \SI{1}{\eV} down to the Fermi energy. In contrast, the valence band (VB) moved only by about \SI{0.7}{\eV}. The different energy shifts of the CB and VB indicate a band gap renormalization of \SI{\approx 0.3}{\eV}.

\newpage

\begin{figure*}[h!]
	\centering
	\includegraphics[width=\columnwidth]{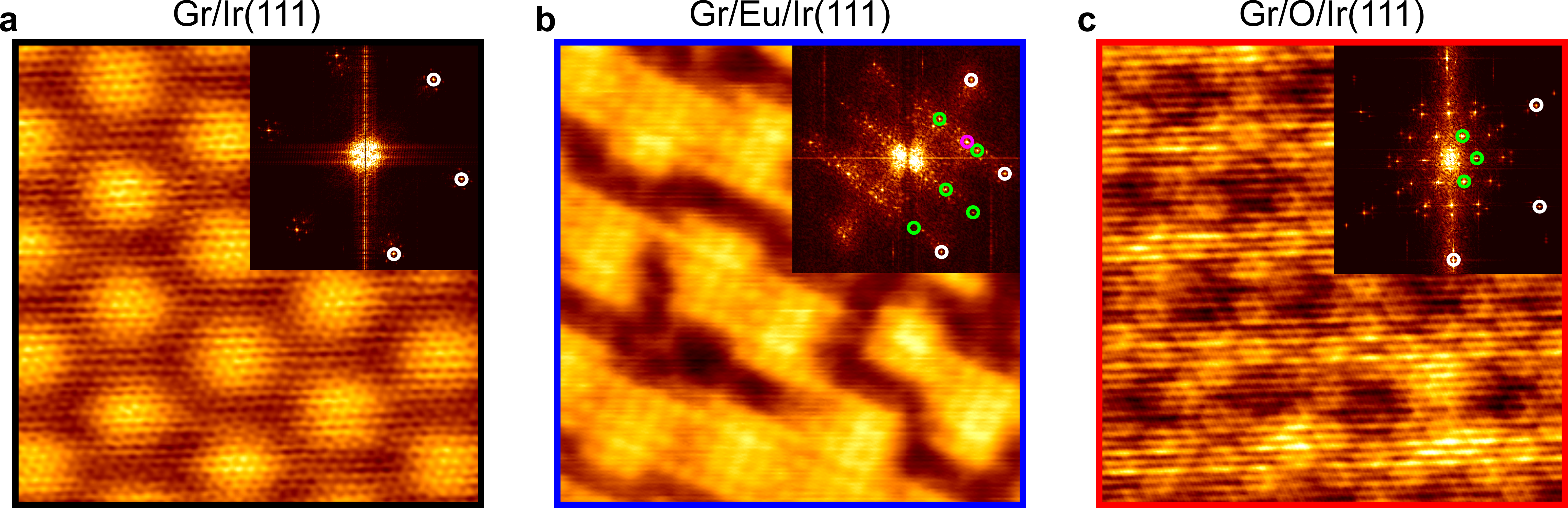}
	\caption{\footnotesize Constant current STM close-up topographs of the Gr substrates: \textbf{a} Gr/Ir(111), \textbf{b} Gr/Eu/Ir(111), \textbf{c} Gr/O/Ir(111). Insets show the Fourier transform (FT) of each image, with spots corresponding to Gr (white circled) and the superstructures (green circled)  highlighted.
			STM parameters ($V$, $I$):
			\textbf{a} $-0.15$\,V, $8.00$\,nA;
			\textbf{b} $0.30$\,V, $0.40$\,nA; 			
			\textbf{c} $0.01$\,V, $0.50$\,nA;
			\textbf{a-d} image size $\SI{10\times10}{\nm^2}$.	
			\label{supp:STMtopo}}
	\end{figure*}

\section*{Fig.~\ref{supp:STMtopo}: Intercalation structures}

In Fig.\,\ref{supp:STMtopo} we identify the graphene (Gr) intercalation structures on the atomic scale. For purpose of comparison, Fig.\,\ref{supp:STMtopo}a shows a topograph of Gr/Ir(111) and, in the inset, the Fourier transform (FT) of the topograph. The Gr lattice and the moir\'{e} with Ir(111), of unit lengths $0.245$\,nm and $2.53$\,nm respectively, are visible in real space. In reciprocal space the brightest outer spots correspond to Gr (white circles).
	
The topograph in Fig.\,\ref{supp:STMtopo}b shows an area of Gr/Eu/Ir. A rectangular lattice covers most of the image, with small trenches of presumably lower (or zero) Eu density in between. The Eu superstructure (green circles) is measured to have unit lengths $(0.426\pm0.07)$\,nm and $(0.495\pm0.06)$\,nm, corresponding to a c$(4\times2)$ with respect to Gr. The superstructure spots also have satellites of the Gr/Ir(111) moir\'{e} unit length, and there are additional spots (pink circle), each resulting from the summation of vectors analogous to multiple scattering in LEED. Eu is known to form various hexagonally symmetric superstructures under Gr on Ir(111) \cite{Foerster2012,Schumacher2013b,Schumacher2014}. The c$(4\times2)$ phase has not been reported previously, but we note that it has precisely the same density as the well-documented $(2\times2)$ phase, namely $25$\percent~with respect to the Gr lattice. Its electronic effect on Gr is also comparable, see Fig.~\ref{supp:figfes}.

In Fig.\,\ref{supp:STMtopo}c the O intercalation structure is seen. It forms a $(2\sqrt3 \times 2\sqrt3)$-R$30\degree$ superstructure with respect to Ir(111), which was previously reported in Ref.~\citenum{Martinez-Galera2016}. It has a density of 0.5 ML of O with respect to the Ir(111) surface.

\newpage

\begin{figure*}[h!]
	\centering
	\includegraphics[width=0.7\columnwidth]{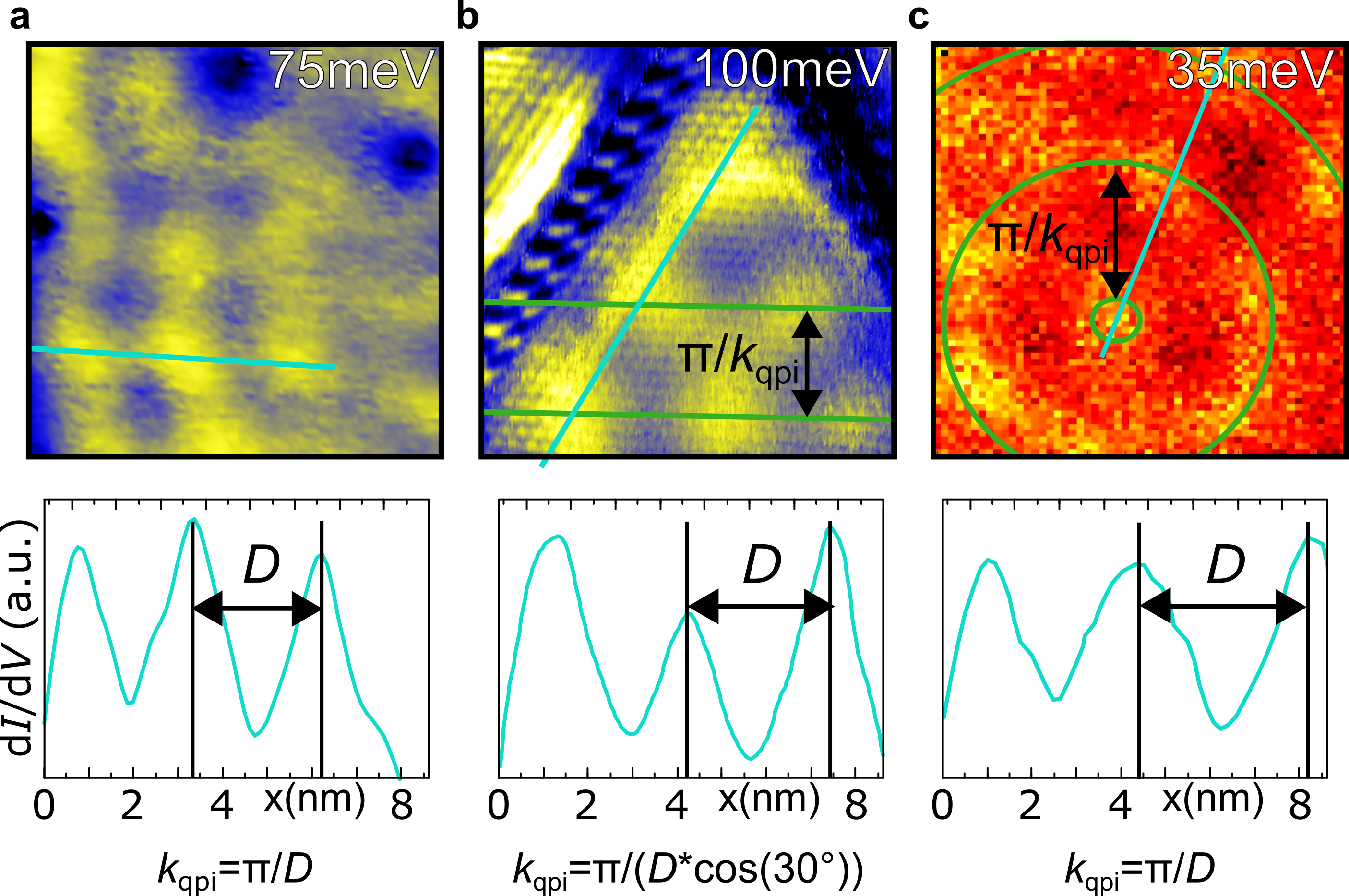}
	\caption{\footnotesize Extraction of the dispersion from STS data of n-shifted \mo. \textbf{a}~Top panel: differential constant current conductance map. Quasiparticle interference patterns around defects are visible. A line profile (cyan line) is taken on maxima normal to a \mo{} island edge. \textbf{b}~Top panel: differential constant current conductance map of a triangular quantum well. A line profile is taken on the maxima parallel to an edge of the well. \textbf{c}~Top panel: constant height STS grid at 35 meV of an  approximately circular quantum well. A line profile is taken along maxima so that the line is normal to the circular wavefronts. \textbf{a-c}~Lower panel: line profile along cyan line in top panel, with below it the relation of maxima along the line profile ($D$) and wavenumber ($k_{\text{qpi}}$).
	STM/STS parameters:
			\textbf{a} $I_\text{st}$=\SI{0.1}{\nA}, image size \SI{11 \times 11}{\nano\meter\squared};
			\textbf{b} $I_\text{st}$=\SI{0.1}{\nA}, image size \SI{9 \times 9}{\nano\meter\squared};
			\textbf{c} $I_\text{st}$=\SI{0.4}{\nA}, image size \SI{9.7 \times 9.7}{\nano\meter\squared}.
			\label{supp:figdis}}
	\end{figure*}

\section*{Fig.~\ref{supp:figdis}: Fitting of the dispersion}

To determine the dispersion of the quasiparticles crossing the Fermi level, the real space wavelength of standing waves due to quasiparticle interference $D$ can be measured. The exact relation between the maxima visible in differential conductance maps and the wavenumber $k_{\text{qpi}}$ depends on the geometry of the scattering area. The most straightforward case is depicted in Fig.~\ref{supp:figdis}a. When quasiparticles scatter from a (relatively) straight edge or boundary, the standing wave normal to the scatterer has a wavenumber $k_{\text{qpi}}= \pi/D$. Note that there are multiple scatters present in the area, which causes the behavior of the quasiparticle interference (QPI) to differ from that of a single wave-train scattering from a 1D defect~\cite{Crommie1993b}.

In the case of a triangular quantum well, as in Fig.~\ref{supp:figdis}b, interference from the three straight edges leads to a more complicated pattern, which can be solved exactly, see Ref.~\citenum{Li1985}. In real space, one can measure the wavefront along a straight edge, which is a factor of $\cos(30\degree)$ larger than the wavelength of the quasiparticles.

We can approximate the area in Fig.~\ref{supp:figdis}c as a circular quantum well, leading to ringlike interference patterns. Exact solutions exist for a circular quantum well, see \textit{e.g.} Ref.~\citenum{Jolie2014}. In real space, one can measure the difference between the radii of the circular interference patterns to obtain $k_{\text{qpi}}$~\cite{Jolie2015}.

\newpage

\begin{figure*}[h!]
	\centering
	\includegraphics[width=0.5\columnwidth]{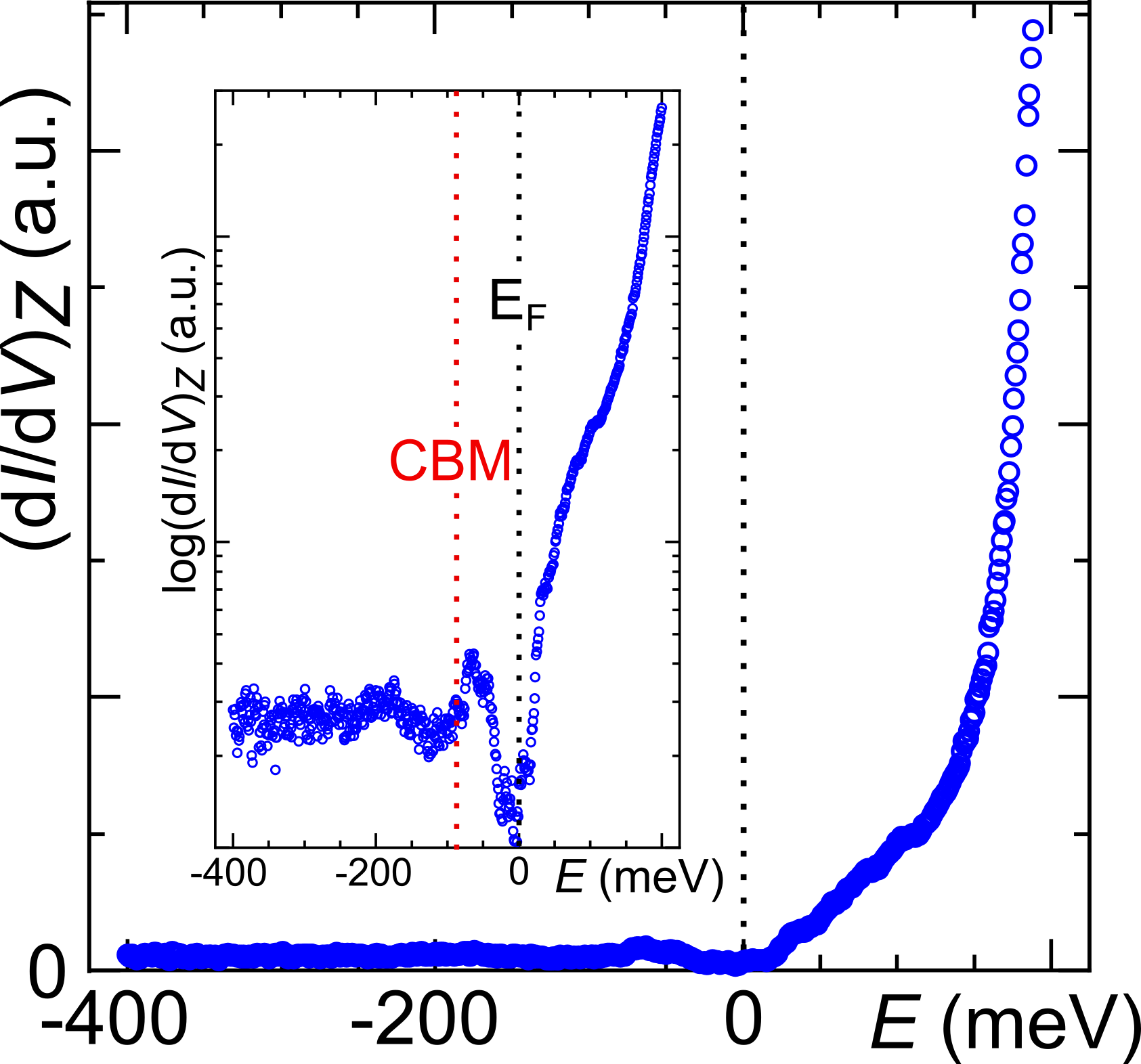}
	\caption{\footnotesize Constant height STS performed close to the surface. In the inset the same data is shown, plotted in a logarithmic scale. The CB minimum (red) and the Fermi level (black) are indicated. $V_\text{st}$= \SI{0.2}{\V}, $I_\text{st}$=\SI{0.5}{\nA}.
			\label{supp:figCBM}}
	\end{figure*}

\section*{Fig.~\ref{supp:figCBM}: Constant height STS of CB minimum}

The CB minimum can also be measured with constant height STS when decreasing the tip-sample distance. We find that the differential conductance decreases towards the Fermi level when stabilizing at \SI{0.2}{\eV} but remains clearly above the noise level well into the occupied states. The signal disappears at \SI{-80}{\milli\eV}, marking the onset of the CB in agreement with our QPI analysis.

We find a pronounced gap around the Fermi level. While the origin of this gap remains unclear at present, we speculate that it is a consequence of phonon-mediated inelastic tunneling. A similar gap has been observed in graphene~\cite{Zhang2008, Natterer2015}.

\newpage

\section*{Table~\ref{supp:dft}: DFT results}

\begin{table}[h]

		\caption{\footnotesize Charge transfers between MoS$_2$, Gr, and dopant (in units of $e$/supercell) calculated from Bader charges. Potential shifts induced over the heterostructures due to charge transfer (in units of $V$). Values for the doped systems are given relative to the pristine system, where it is $\SI{0.249}{\V}$. $E_\text{D}-E_\text{F}$ is the position of Dirac-point with respect to the Fermi level. The concentration of dopants is indicated as Eu$x$ and O$x$, with $x$ the number of dopant atoms per \mo{} supercell.
}
	\label{supp:dft}
	\begin{tabularx}{0.82\textwidth}{c|EEEEE}
	 & \multicolumn{1}{Y}{$\text{~~~~~~MoS}_2$} & \multicolumn{1}{Y}{~~~~~Gr} & \multicolumn{1}{Y}{~~~~~~dopant} & \multicolumn{1}{Y}{~~~~~~$\Delta V$}&\multicolumn{1}{Y}{~~~$E_\text{D} - E_\text{F}$}\\
	\hline
	MoS$_2$/Gr      & 0.042 & -0.042 &        &  0     & 0 \\
MoS$_2$/Gr/Eu1  & 0.204 &  0.615 & -0.820 &  0.934 & -0.66 \\
MoS$_2$/Gr/Eu2  & 0.265 &  1.388 & -1.653 &  1.119 & -0.99 \\
MoS$_2$/Gr/O1   & 0.041 & -0.352 &  0.311 & -0.612 & 0.34 \\
MoS$_2$/Gr/O2   & 0.040 & -0.535 &  0.494 & -0.940 & 0.58 \\
MoS$_2$/Gr/O3   & 0.040 & -0.703 &  0.663 & -1.241 & 0.61 \\

	\end{tabularx}

\end{table}

\newpage

\begin{figure*}[h!]
		\centering
		\includegraphics[width=0.7\textwidth]{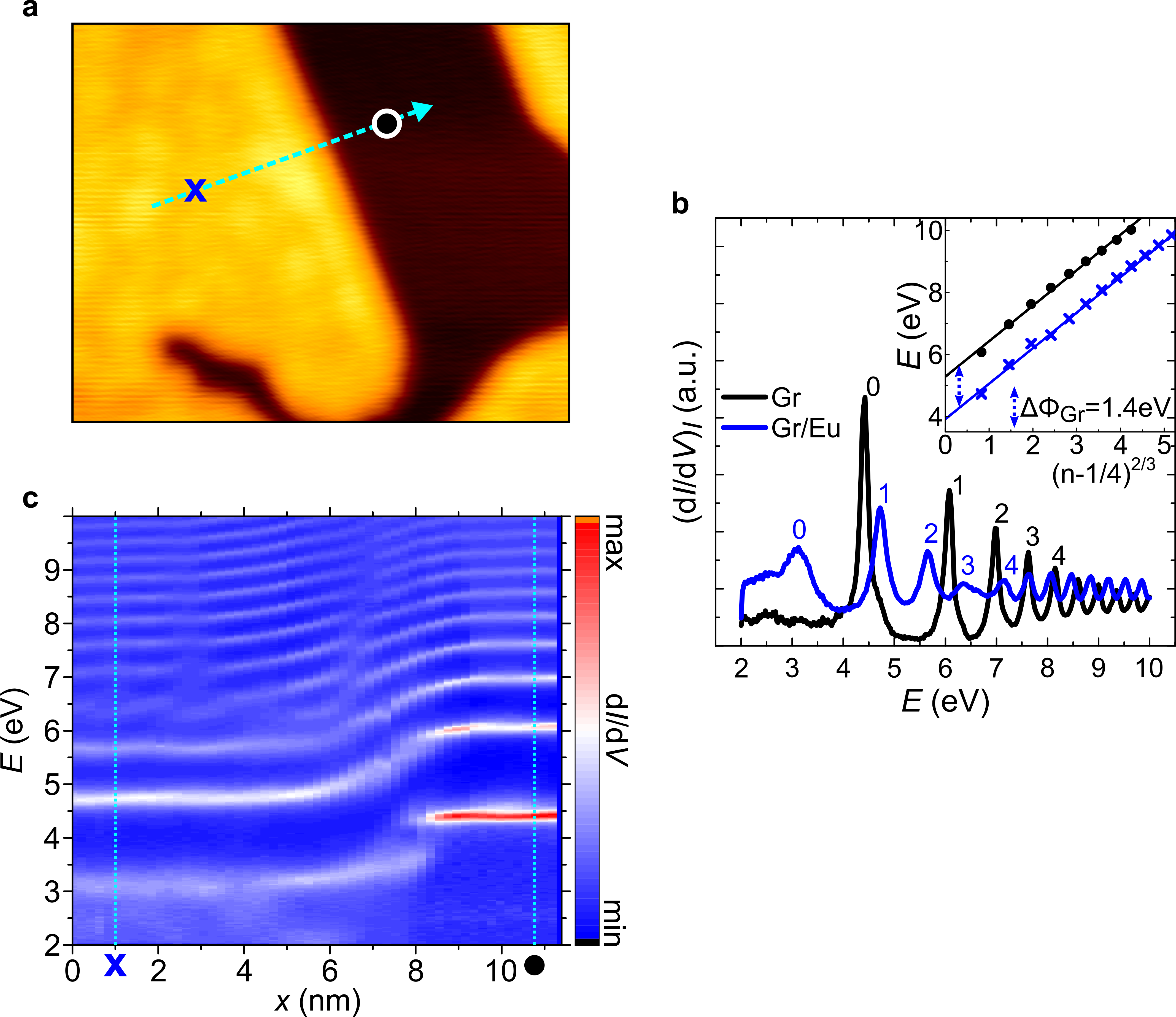}
		\caption{\footnotesize Field emission resonance (FER) spectroscopy on n-doped Gr. \textbf{a} STM topograph showing region of intercalated Gr (bright) and unintercalated Gr (dark). \textbf{b} FER spectroscopy recorded on Gr/Ir(111) (black) and Gr/Eu/Ir(111) (blue) at the locations marked in \textbf{a}. Resonant states are numbered $n$, and in the inset (right) plotted $E$ against $(n-\frac{1}{4})^\frac{2}{3}$, see Ref.\,\citenum{Lin2007}. The energy shift in the corresponding lines of best fit gives the difference in Gr work function $\Delta\Phi_{\text{Gr}}=(-1.4\pm0.1)$\,eV. \textbf{c} Constant current linescan along the cyan line shown in \textbf{a}.
		STM and STS parameters ($V$, $I$):
		\textbf{a} $20\times16$\,nm$^2$, $1.0$\,V, $0.04$\,nA;
		\textbf{b, c} FER spectroscopy stabilization voltage $V_\text{st}=2$\,V, $I_\text{st}=0.1$\,nA.
			\label{supp:figfes}}
	\end{figure*}

\section*{Fig~\ref{supp:figfes}: Field emission resonances on n-doped Gr}

Since the c$(4\times2)$ phase of Eu under Gr has not been observed before, we checked the doping level of Gr using field emission resonance (FER) spectroscopy. FER allows us to determine the change in the Gr work function due to the Eu intercalation layer. In Fig.\,\ref{supp:figfes}a an area of partly intercalated Gr is shown. The bright patches are Eu-intercalated Gr and the dark regions bare Gr. Constant current STS point spectra were taken at the cross and circle positions in Fig.\,\ref{supp:figfes}a. The spectra are shown in Fig.\,\ref{supp:figfes}b. The d$I$/d$V$ peaks are resonant field emission states. In accordance with the method described by Lin \textit{et al.}~\cite{Lin2007}, each state can be assigned an order $n$ and an approximately linear energy shift between same-order states observed, see inset of Fig.\,\ref{supp:figfes}b. This energy shift indicates a work function reduction $\Delta\Phi_{\text{Gr}}=(-1.4\pm0.1)$\,eV on Eu-intercalated Gr. Assuming a rigid shift of the Dirac cone, this would mean that the Dirac point has shifted to $-1.3$\,eV, compared to its value of $0.1$\,eV for Gr/Ir(111) \cite{Pletikosic2009}. This is in line with the ARPES-measured Dirac point of $-1.36$\,eV for Gr/$(2\times2)$-Eu/Ir(111) \cite{Schumacher2014}, and represents strong n-doping of Gr. 

Looking at Fig.\,\ref{supp:figfes}c, where a constant current STS linescan, taken along the line in Fig.\,\ref{supp:figfes}a, is depicted, we find that the Eu-intercalated islands are near-uniformly doped, as the FER states show little change when going across the doped Gr region.

\newpage

\section*{Table~\ref{supp:tbkf}: Charging of MTBs}

\begin{table}[h]
		\begin{tabularx}{0.6\textwidth}{c|EEE}
			\mo~on & \multicolumn{1}{Y}{$k_\text{F}\,(\pi/a)$~} & \multicolumn{1}{Y}{~~$\lambda\,(e/a)$} & \multicolumn{1}{Y}{$\Delta E_{\text{F}}$\,(eV)}\\ 
			\hline
			freestanding (DFT) \cite{Murray2020} & $~0.63$  & $~~-0.74$ & -\\ 
			Gr (DFT) \cite{Murray2020} & $0.54$ & $-0.92$ & -\\
			Gr/Ir (STM) & $0.50$  & $-1.00$ & 0\\ 
			Gr/Eu/Ir (STM) & $0.37$  & $-1.26$ &+0.33\\ 
			Gr/O/Ir (STM) & $0.58$ & $-0.84$ &-0.19\\
		\end{tabularx}
		\caption{\footnotesize Fermi wavevectors $k_\text{F}$ of the 4|4E MTB band in ML-\mo{} on different substrates, and the associated line charge on the MTB, $\lambda = -\frac{2e}{\pi}(1-k_\text{F})$. See Ref.~\citenum{Murray2020} for more details. The experimentally determined shifts of the Fermi level due to doping of the Gr substrate are given with respect to the pristine (undoped) case.
	}
	\label{supp:tbkf}
\end{table}

In previous work, DFT showed that a 4|4E MTB in a \textit{freestanding} \mo{} layer is approximately charge neutral, with a line charge $\lambda_{\text{MTB}}^{\text{band}}$ of about the same magnitude (but opposite polarity) as the polarization charge of the MTB $\lambda_{\text{MTB}}^{\text{pol}}$. This leads to $\lambda_{\text{MTB}}=\lambda_{\text{MTB}}^{\text{band}} - \lambda_{\text{MTB}}^{\text{pol}} \approx 0$. For details see Ref.~\citenum{Murray2020}. Placing the \mo{} on Gr caused an increased MTB band filling due to Gr electron donation; comparing Fermi wavevectors before and after showed a changed line charge density of $\Delta \lambda = -\frac{2e}{\pi}(k_\text{F}-k'_\text{F}) = - 0.18\,e/a$~\cite{Murray2020}, see Table~\ref{supp:tbkf}. In STM, a slightly larger value of $- 0.26\,e/a$ was obtained, as the observed Fermi wavevector was $k_{\text{F}}\approx(0.5)\,\pi/a$, instead of the DFT-predicted $k_{\text{F}}= (0.54)\,\pi/a$.

Using STM, we can similarly assess the change in band filling due to the Eu and O interlayer, relative to the unintercalated system. We find $\Delta \lambda_{Eu} = -\frac{2e}{\pi}(k_\text{F}^{\text{Gr}}-k_\text{F}^{\text{Eu}}) = - 0.26\,e/a$  and $\Delta \lambda_O = -\frac{2e}{\pi}(k_\text{F}^{\text{Gr}}-k_\text{F}^{\text{O}}) = + 0.16\,e/a$, where $k_\text{F}^{\text{Gr}}$ is the unintercalated experimental system. 

In addition, we can use the shifts in $k_\text{F}$ to estimate the shift in the Fermi level $E_\text{F}$ of the MTB, relative to the pristine Gr/Ir(111) substrate. This is done in Fig.~6(d) of the main text, where overlapping $k$ states between the three samples are used to map the dispersion of the 1D band. The results for the Fermi level of the MTB are shown in the last column of Table~\ref{supp:tbkf}.

\begin{figure*}[h!]
		\centering
		\includegraphics[width=0.7\textwidth]{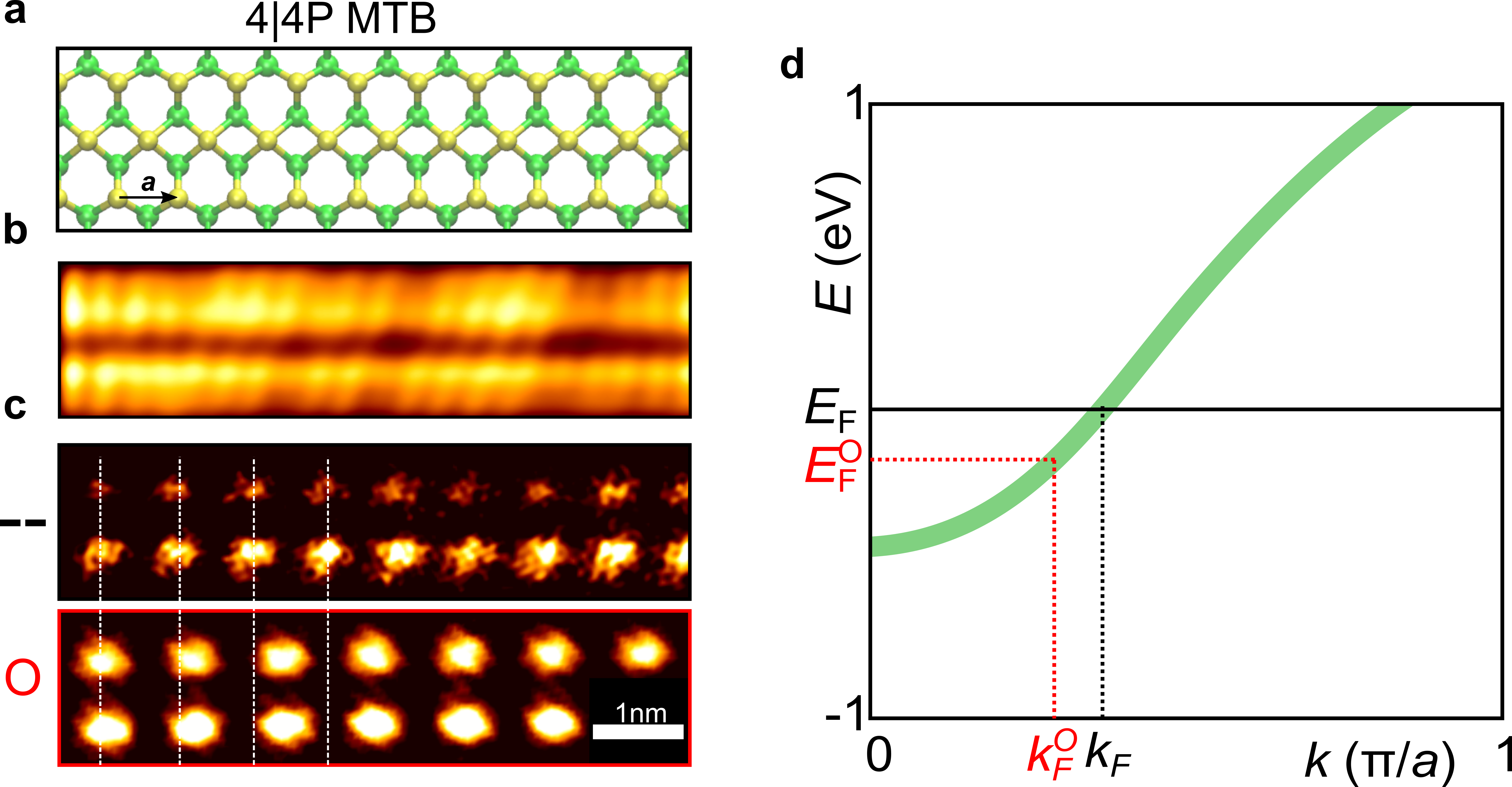}
		\caption{\footnotesize Confined states in 4|4P MTBs in ML-\mo{} on different substrates. 
		\textbf{a}~Ball-and-stick model of 4|4P MTB. Mo atoms are navy, S atoms yellow (top layer) or dark yellow (bottom layer). A \mo{} lattice vector is indicated.
		\textbf{b}~Constant current STM topograph of 4|4P MTB in \mo{} on Gr/Ir. 
		\textbf{c}~Constant height STM of 4|4P MTBs in pristine (--) \mo{} and in \mo{} on p-doped Gr (O). Vertical white lines are a guide to the eye. No 4|4P MTBs were found in \mo{} on n-doped Gr. 
		\textbf{d}~Dispersion of the hole-like 1D band of the 4|4P MTB. Based on DFT calculations from Ref.~\citenum{Murray2020}. Fermi level and $k$-vector based on experimental data. 
		STM/STS parameters: \textbf{b}~$V$=\SI{0.9}{\V}, $I$=\SI{0.02}{\nA}; \textbf{c}~top panel $V$=\SI{0.1}{\V}, $I_\text{st}$=\SI{0.1}{\nA}; bottom panel $V$=\SI{0.04}{\V}, $I_\text{st}$=\SI{0.1}{\nA}. \SI{1}{\nm} scale bar shown.
			\label{supp:fig44p}}
	\end{figure*}

\newpage
\section*{Fig.~\ref{supp:fig44p}: 4|4P MTB}

\newpage

	\bibliographystyle{naturemag}
	\bibliography{./library}